\documentclass[sigconf]{acmart}
\usepackage[T1]{fontenc}
\usepackage{aecompl}
\usepackage{booktabs}
\usepackage{amsmath,amsfonts}

\usepackage{algorithmic}
\usepackage{graphicx}
\usepackage{subfigure}
\usepackage{caption}
\usepackage{textcomp}
\usepackage{tabularx}
\usepackage[ruled,vlined]{algorithm2e}
\usepackage[table,xcdraw]{xcolor}

\usepackage{framed}
\usepackage{bm}
\usepackage{soul}
\usepackage{subcaption}
\usepackage{enumerate}
\usepackage{tcolorbox}
\usepackage{xcolor}
\usepackage{arydshln}
\usepackage{multirow}
\usepackage{hyperref}
\usepackage[export]{adjustbox}
\usepackage{enumitem}
\usepackage{microtype}
\usepackage{balance}
\usepackage{array}
\usepackage{makecell}
\usepackage{todonotes}
\newcolumntype{C}[1]{>{\centering\arraybackslash}m{#1}} 
\newcolumntype{Y}{>{\raggedright\arraybackslash}X} 

\setitemize[1]{itemsep=0pt,partopsep=0pt,parsep=\parskip,topsep=5pt}

\def\BibTeX{{\rm B\kern-.05em{\sc i\kern-.025em b}\kern-.08em
    T\kern-.1667em\lower.7ex\hbox{E}\kern-.125emX}}

\newcommand{\hz}[1]{\textcolor{black}{#1}}

\newcommand{\tool}[1]{\textsc{#1}\xspace}

\newcommand{\nonthinkmode}{\tool{Non-Think}}
\newcommand{\thinkmode}{\tool{Think}}
\newcommand{\mydashline}{\noalign{\vskip 0.2ex}\cdashline{2-17}\noalign{\vskip 0.3ex}}

\AtBeginDocument{%
  \providecommand\BibTeX{{%
    \normalfont B\kern-0.5em{\scshape i\kern-0.25em b}\kern-0.8em\TeX}}}

\copyrightyear{2026}
\setcopyright{rightsretained}
\acmConference[SIGIR '26]{Proceedings of the 49th International ACM SIGIR Conference on Research and Development in Information Retrieval}{July 20--24, 2026}{Melbourne, VIC, Australia}
\acmBooktitle{Proceedings of the 49th International ACM SIGIR Conference on Research and Development in Information Retrieval (SIGIR '26), July 20--24, 2026, Melbourne, VIC, Australia}

\begin{document}

\title{Think When Needed: Model-Aware Reasoning Routing for LLM-based Ranking}

\author{Huizhong Guo}
\orcid{0009-0004-0011-8612}
\authornote{Equal Contribution}
\affiliation{%
  \institution{Zhejiang University}
  \city{Hangzhou}
  \country{China}
}
\email{huiz_g@zju.edu.cn}

\author{Tianjun Wei}
\orcid{0000-0001-7311-7101}
\authornotemark[1]
\affiliation{%
  \institution{Nanyang Technological University}
  \country{Singapore}
}
\email{tjwei2-c@my.cityu.edu.hk}

\author{Dongxia Wang}
\orcid{0000-0001-9812-3911}
\authornote{Corresponding author.}
\affiliation{%
  \institution{Zhejiang University}
  \city{Hangzhou}
  \country{China}
}
\email{dxwang@zju.edu.cn}

\author{Yingpeng Du}
\orcid{0000-0001-9881-7171}
\affiliation{%
  \institution{Nanyang Technological University}
  \country{Singapore}
}
\email{yingpeng.du@ntu.edu.sg}

\author{Ziyan Wang}
\orcid{0000-0001-5463-7099}
\affiliation{%
  \institution{Nanyang Technological University}
  \country{Singapore}
}
\email{wang1753@e.ntu.edu.sg}

\author{Jie Zhang}
\orcid{0000-0001-8996-7581}
\affiliation{%
  \institution{Nanyang Technological University}
  \country{Singapore}
}
\email{zhangj@ntu.edu.sg}

\author{Zhu Sun}
\authornotemark[2]
\orcid{0000-0002-3350-7022}
\affiliation{%
  \institution{Singapore University of Technology and Design}
  \country{Singapore}
}
\email{sunzhuntu@gmail.com}
\renewcommand{\shortauthors}{Huizhong Guo et al.}

\begin{abstract}
Large language models (LLMs) are increasingly applied to ranking tasks in retrieval and recommendation.
Although reasoning prompting can enhance ranking utility, our preliminary exploration reveals that its benefits are inconsistent and come at a substantial computational cost, suggesting that \textit{when to reason} is as crucial as \textit{how to reason}.
To address this issue, we propose a reasoning routing framework that employs a lightweight, plug-and-play router head to decide whether to use direct inference (\nonthinkmode) or reasoning (\thinkmode) for each instance before generation. 
The router head relies solely on pre-generation signals: i) compact \textbf{ranking-aware features} (e.g., candidate dispersion) and ii) \textbf{model-aware difficulty signals} derived from a diagnostic checklist reflecting the model's estimated need for reasoning.
By leveraging these features before generation, the router outputs a controllable token that determines whether to apply the \thinkmode mode.
Furthermore, the router can adaptively select its operating policy along the \textbf{validation Pareto frontier} at deployment time, enabling dynamic allocation of computational resources toward instances most likely to benefit from \thinkmode under varying system constraints.
Experiments on three public ranking datasets with different scales of open-source LLMs show consistent improvements in ranking utility with reduced token consumption (e.g., \textbf{+6.3\%} NDCG@10 with \textbf{–49.5\%} tokens on MovieLens with Qwen3-4B), demonstrating reasoning routing as a practical solution to the accuracy-efficiency trade-off.
\end{abstract}

\ccsdesc[500]{Information systems~Recommender systems}
\ccsdesc[500]{Information systems~Language models}

\keywords{List-wise Ranking, Large Language Model, Adaptive Reasoning}
\maketitle

\section{Introduction}
Large language models (LLMs) are increasingly applied to ranking tasks, including information retrieval~\cite{sun-etal-2023-chatgpt, liu2024information} and personalized recommendation~\cite{hong2025llm, yin2025unleash}.
Unlike traditional ranking models, LLMs can integrate heterogeneous signals within a unified generative framework and reason over long, complex inputs.
Moreover, recent studies demonstrate that prompting LLMs with reasoning (\thinkmode\footnote{In this work, we use \nonthinkmode and \thinkmode to denote the mode that outputs directly without a reasoning process and the mode that employs reasoning, respectively.})~\cite{wei2022chain} can further enhance ranking utility, yielding more accurate and fine-grained item ordering~\cite{sun2024large}.
These improvements are attributed to the model's capacity to resolve subtle relevance distinctions through multi-step reasoning.

\begin{figure}[t]
  \setlength{\abovecaptionskip}{2pt}
  \centering
    \includegraphics[width=\linewidth]{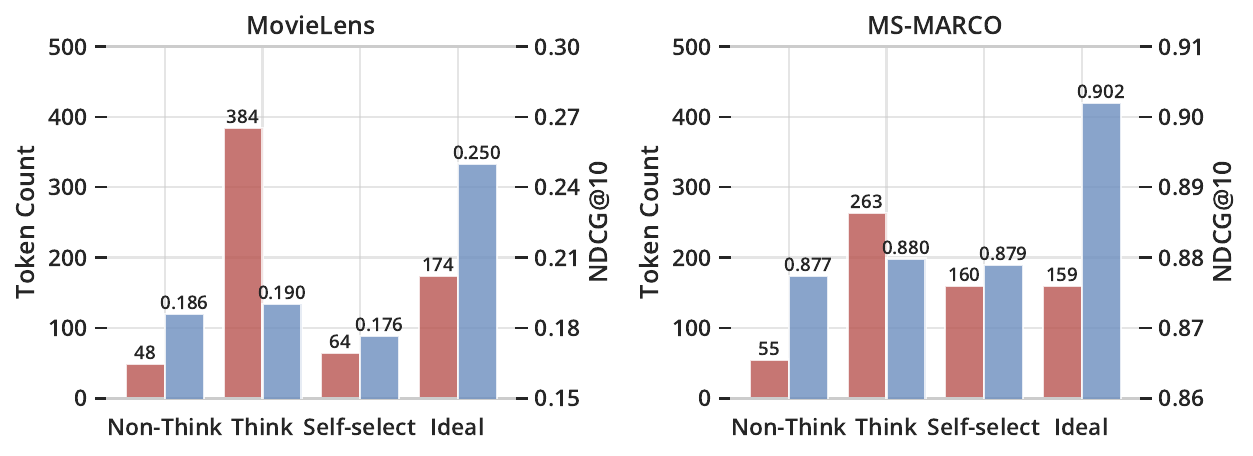}
  \caption{\textbf{Accuracy–efficiency trade-offs across reasoning modes.} Left axis shows token count; right axis shows NDCG@10. Modes: \nonthinkmode, \thinkmode, \textsc{Self-Select} (the LLM chooses the mode per instance), and \textsc{Ideal} (an oracle selects the better mode for each instance).}
  \label{fig:exp_motivation}
  \vspace{-0.4cm}
\end{figure}

However, ranking systems typically operate as critical components within larger retrieval or recommendation pipelines, where response latency is a primary concern~\cite{parry2024top}.
As shown in Figure~\ref{fig:exp_motivation},
\thinkmode consumes roughly \textbf{8 times} as many tokens as \nonthinkmode, prohibiting its use in latency-sensitive systems.
Furthermore, our empirical analysis reveals that the benefits of \thinkmode are not universal, as \nonthinkmode surpasses it in a substantial portion of instances.\footnote{The detailed analysis of these findings is presented in Section~\ref{sec:motivation}.}
\hz{In many such cases, the reasoning process can be hindered by ambiguous or insufficient signals, causing the model to engage in prolonged or ineffective reasoning loops that ultimately lead to task failure rather than improvement.}
Therefore, an effective ranking system should not rely on simply increasing reasoning effort to improve ranking accuracy; instead, it should adaptively choose to skip reasoning when appropriate. 

This paradigm of deciding \textit{when to reason} is known as \textit{adaptive reasoning}~\cite{aggarwal2025l1,wu2025arm}, and represents a promising direction for achieving practical accuracy-efficiency trade-offs.
Existing research has primarily focused on general reasoning scenarios, such as mathematical problem-solving and code generation~\cite{chen2024toward, wu2025arm}. 
These domains are characterized by \textbf{well-defined problem formulations}, \textbf{model-independent measures of difficulty}, and \textbf{directly verifiable answers}, enabling methods to adjust reasoning depth (e.g., number of steps)~\cite{wan2025adapthink, hou2025thinkprune} based on explicit correctness feedback.
In contrast, adaptive reasoning for ranking systems faces several unique challenges that distinguish it from prior work:\

\begin{itemize}[leftmargin=8pt]
\item[$\blacklozenge$] \textbf{First, difficulty in ranking lacks an explicit and objective measure}.
Ranking complexity arises from multiple intertwined factors in the input, such as inter-candidate relationships and query-candidate associations.
These factors interact in nontrivial ways, making the overall task difficulty challenging to quantify directly.
Moreover, the results of \textsc{Self-Select} show that current LLMs struggle to infer such difficulty from input, failing to distinguish between straightforward and challenging instances.

\item[$\blacklozenge$] \hz{\textbf{Second, the benefit of reasoning in ranking varies substantially across instances and is largely model-dependent.}
Ranking tasks typically involve a substantial proportion of noisy or underspecified instances.
Consequently, for many such cases, allocating additional reasoning effort yields no measurable benefit, as the available preference signal is intrinsically weak or ambiguous.
Furthermore, the perceived difficulty of a ranking instance may vary across LLMs (see Section~\ref{sec:motivation}), underscoring the limitations of a universal reasoning router for all LLMs.}

\item[$\blacklozenge$] \hz{\textbf{Third, ranking systems are inherently cost-sensitive}. 
Unlike logic-centric tasks such as mathematics or code, where maximizing per-instance accuracy is often the dominant objective, ranking systems must jointly optimize ranking utility and inference cost. 
In large-scale retrieval and recommendation settings, token budgets directly translate into system latency, making computational cost a first-class concern~\cite{fu2024efficient, podolak2025beyond}. 
Accordingly, reasoning resources should be allocated selectively based on the expected utility of each instance.
}
\end{itemize}

To address the limitations identified above, we propose a lightweight, plug-and-play framework that tackles the fundamental challenge of determining \textit{when an LLM should adopt complex reasoning} to maximize ranking utility under computational constraints.
\hz{Our core idea is to design a \textbf{\textit{lightweight, model-aware reasoning router}} that can be seamlessly integrated into LLM-based ranking pipelines to assist mode selection.
The router is trained to estimate, at the instance level, the expected utility gain of applying \thinkmode over \nonthinkmode. 
First, we construct a compact set of \textit{\textbf{ranking-aware features}} to characterize the intrinsic structural complexity of a ranking instance. 
These features quantify properties such as candidate dispersion and context--candidate alignment, providing an explicit signal of task difficulty prior to generation. 
Second, we introduce a checklist-based probing mechanism to capture \textit{\textbf{model-aware difficulty signals}} by assessing the LLM’s own perception of instance difficulty before inference. 
The resulting signals, derived from the model’s probabilistic responses, reflect internal uncertainty and potential reasoning failure modes, and are incorporated as a complementary feature set without modifying the backbone model or its decoding process.
Once trained, the reasoning router enables flexible, instance-level reasoning allocation under practical computational constraints.
At inference time, this router utilizes the \textit{\textbf{validation Pareto frontier}} to select among flexible operating policies.}
This enables the system, under given computational constraints, to allocate \thinkmode only to instances with the highest predicted benefit, achieving an optimal trade-off between accuracy and efficiency.

In summary, our proposed framework offers a unified and efficient approach to adaptive reasoning for ranking tasks, enabling dynamic allocation of reasoning effort and achieving consistently better accuracy–efficiency trade-offs.
Building on this foundation, our work makes the following key contributions:
\begin{itemize}[leftmargin=0.4cm]
\item We present the first formulation of \textbf{reasoning routing for ranking tasks}, showing that long reasoning is \emph{not universally beneficial} for LLM-based ranking and motivating a \emph{per-instance} mode router to balance accuracy and efficiency.
\item We propose a \textbf{lightweight, model-aware router} that adaptively chooses between \nonthinkmode and \thinkmode for each instance. The router leverages \textit{ranking-aware features} and \textit{model-aware difficulty signals} to make decisions \emph{before} generation, without modifying the backbone architecture or decoding process.
\item We conduct extensive experiments across three public datasets and different scales of open-source LLMs, demonstrating consistent accuracy gains with lower token usage (e.g., \textbf{+6.3\%} NDCG@10 with \textbf{–49.5\%} tokens on MovieLens with Qwen3-4B).\footnote{Our code is available at: \url{https://github.com/Grey-z/reasoning_router}.}
\end{itemize}

\begin{figure}[ht]
    \setlength{\abovecaptionskip}{1pt}
    \centering
    \subfigure[Instance-wise Mode Superiority.]{
        \centering
    \includegraphics[width=0.56\linewidth]{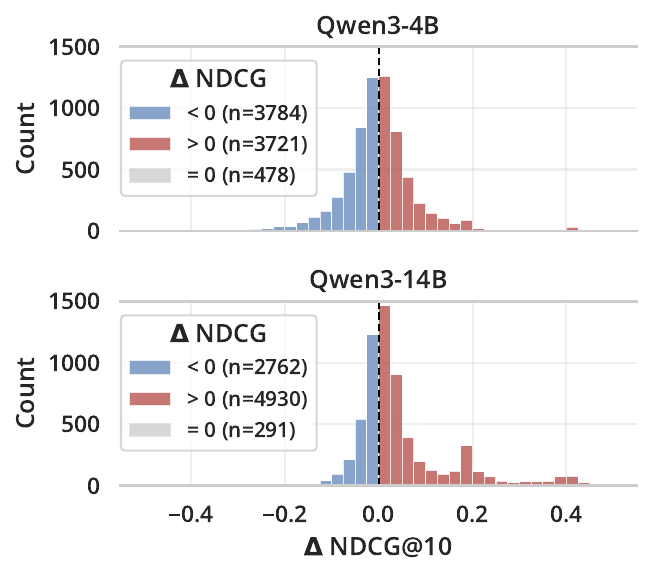}
        \label{fig:pie_case}
    }
    \hspace{-0.4cm}
    \subfigure[Model Consistency]{
        \centering
        \includegraphics[width=0.41\linewidth]{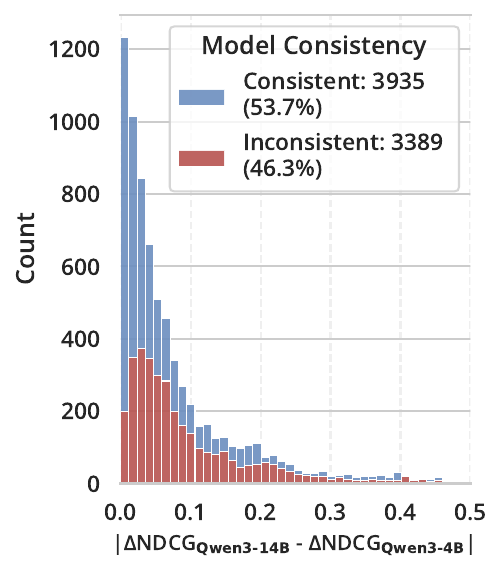}
      \label{fig:model_consistency}
    }
      \caption{Performance with different reasoning modes on MS-MARCO dataset: (a) Difference of NDCG@10 between \thinkmode and \nonthinkmode (denoted by $\Delta \text{NDCG}$). (b) Absolute difference of $\Delta \text{NDCG}$ between Qwen3-14B and Qwen3-4B. } 
  \label{fig:selective-reasoning}
  \vspace{-0.2cm}
\end{figure}

\section{Related Works}
\textbf{LLM for Ranking Task.}
Recent studies have explored leveraging large language models (LLMs) as reasoning-based rankers for both IR~\cite{dong2025understand} and Rec tasks~\cite{wang2025kgbridge, guo2024configurable}.
In the IR task, LLMs have been used to generate and reason over candidate justifications, improving document relevance estimation~\cite{sun-etal-2023-chatgpt, liu2024information}.
Reasoning prompting~\cite{wei2022chain, sun2024large} enables LLMs to articulate intermediate relevance reasoning, enhancing transparency and interpretability.
For example, \citet{sun2024large} shows that a designed reasoning template improves fine-grained ranking by helping models distinguish near-tied documents.
In the Rec task, reasoning has been introduced to explain and refine user–item matching~\cite{wang2025re2llm, yin2025unleash}, where multi-step inference captures user intent and contextual nuances but also increases latency and variance~\cite{zhuang2024setwise}.
Overall, these works demonstrate that reasoning can enhance LLM-based ranking quality but remains computationally expensive.
Our study focuses on adaptive reasoning, learning when reasoning is beneficial, and designing a model-aware router that selectively invokes reasoning to balance ranking accuracy and efficiency.

\textbf{Adaptive Reasoning.}
Recent research on reasoning efficiency has predominantly centered on single-output tasks such as mathematical reasoning and code generation, where task difficulty can be explicitly defined by correctness and test-time computation can be directly traded for accuracy~\cite{chen2024toward, wu2025arm}.
A major line of work optimizes reasoning length by constraining generation during training or inference.
For instance, multi-format supervision during pre-training~\cite{wu2025arm, sudualformer} and reinforcement-learning objectives with explicit length penalties~\cite{wan2025adapthink, hou2025thinkprune} both encourage concise reasoning traces without degrading correctness.
Other studies replace token-level reasoning with latent optimization~\cite{hao2024training, tang2025think}, representing intermediate reasoning steps as continuous vectors to reduce textual redundancy.
However, these methods target question-answer settings rather than listwise tasks such as IR and Rec~\cite{liu2025reasonrank}, where reasoning must integrate signals across candidates.
In contrast, we study adaptive routing for ranking, deciding per instance whether to apply reasoning, rather than merely reducing its depth.

\section{Motivation}
\subsection{Preliminaries}
In this work, we study two common ranking tasks—\emph{information retrieval} (IR) and \emph{personalized recommendation} (Rec).
Both can be formalized as selecting and ordering the top-\(K\) items from a candidate set given a task instruction and a task-specific context.
We assume an upstream retriever or sampler provides the candidate set.
Therefore, our focus is on \emph{how an LLM scores and ranks these candidates}, not on candidate generation.

\emph{Problem Setup.}
Let \(P\) denote the task instruction prompt, and let \(\mathcal{I}=\{i_1,\dots,i_n\}\) be the candidate set.
Given context \(C\) and target length \(K \le n\), the LLM \(f_{\theta}\) assigns a score \(s_{\theta}(i\mid P, C, \mathcal{I}) \in \mathbb{R}\) for each candidate item \(i\) and returns a top-\(K\) ranked list
\begin{equation}
L^{K} \;=\; \tau_{K}\!\left(\{\, s_{\theta}(i \mid P, C, \mathcal{I}) : i \in \mathcal{I} \,\}\right), 
\end{equation}
where \(\tau_{K}(\cdot)\) returns the \(K\) highest-scoring items.

For the IR task, the context is a search query \(Q\) (i.e., \(C=Q\)) and candidates are documents or passages.
Given \((P, Q, \mathcal{I})\), the LLM produces a list with items ordered by predicted relevance to \(Q\).
For the Rec task, the context is a user's interaction history \(H\) (i.e., \(C=H\)) and candidates are items for recommendation (e.g., movies).

\begin{figure}[t]
\setlength{\abovecaptionskip}{0pt}
\setlength{\belowcaptionskip}{2pt} 
\begin{tcolorbox}[width=\linewidth]
\scriptsize
You are a ranking assistant. Your task is to \ldots <TASK DESCRIPTION>

\textcolor{blue}{<QUERY> <CANDIDATES>}

When generating the ranking:

- <CONSTRAINT>

Depending on the \textbf{complexity of the ranking task}, you may choose whether to include a
reasoning process:

\textcolor{red}{
\:\:- If the case is \textbf{simple and straightforward}, directly output the result without reasoning.}

\textcolor{red}{\:\:- If the case is \textbf{ambiguous or difficult}, include a reasoning process to justify your
ranking. The reasoning must be wrapped inside <thought> </thought> tags.}

Your final output must \textbf{strictly follow the required format}.

\#\# Output Format

<output>Ranking result: [ITEM IDS IN ORDER, SEPARATED BY COMMA]</output>
\end{tcolorbox}
\caption{Prompt template for reasoning routing.}
\label{fig:prompt_all_2}
\vspace{-0.3cm}
\end{figure}

\subsection{Diagnostic Experiments and Key Findings}
\label{sec:motivation}
In this section, we conduct controlled comparative experiments to quantify the performance of different reasoning modes on ranking tasks.
We adopt the open-source LLMs (\textbf{Qwen3-4B/8B/14B}~\cite{qwen3technicalreport}) as the backbone models and evaluate two ranking tasks: information retrieval on \textbf{MS-MARCO}~\cite{pradeep2023rankzephyr}, and personalized recommendation on \textbf{MovieLens}~\cite{harper2015movielens} (movies) and \textbf{Amazon-VG}~\cite{ni2019justifying} (video games).
For each query or user history, we run two inference modes, \nonthinkmode and \thinkmode, using the same prompt, with the mode determined solely by a prefix token: \textit{<thought>} triggers reasoning process, and \textit{<output>} indicates generating results without reasoning (highlighted in red in Figure~\ref{fig:prompt_all_2}).
We get two ranked lists for each instance, which we then evaluate along two dimensions: (i) \emph{ranking utility} (e.g., NDCG) and (ii) \emph{inference cost} (tokens).\footnote{Complete experimental settings and additional model results are provided in Section~\ref{sec:exp}.}

First, Figure~\ref{fig:exp_motivation} reports dual-axis bar charts comparing four modes: \nonthinkmode (with fixed prefix \textit{<output>}), \thinkmode (with fixed prefix \textit{<thought>}), \textsc{Self-select} (let the LLM choose its reasoning mode), and \textsc{Ideal} (an idealized setting in which the optimal mode is selected for each instance). 
The left axis shows token usage, and the right axis shows ranking utility (NDCG@10).
For example, on MovieLens, \thinkmode improves NDCG by \textbf{2.1}\% over \nonthinkmode, yet increases token consumption by approximately \textbf{7} times.
The results on Amazon-VG and MS-MARCO exhibit a similar trend. 
These findings indicate that \textit{\textbf{\thinkmode improves accuracy on average but at a substantial computational cost}}.
Second, Figure~\ref{fig:pie_case} shows the distribution of instances where each mode achieves the highest accuracy: \nonthinkmode outperforms \thinkmode in 35\% of cases (on MS-MARCO with Qwen3-14B), indicating that \textbf{\textit{\thinkmode is not universally superior}}.
This outcome is plausible because, in long structured inputs, multi-step reasoning may dilute key evidence, amplify early errors, and distract from fine-grained cues that separate closely ranked candidates.
In addition, 3.6\% of instances yield zero accuracy under both modes, suggesting these instances are likely beyond the model's capacity.
For such instances, invoking reasoning is unnecessary, as it increases cost without improving ranking utility.
Third, the results of \textsc{Self-select} show that allowing the LLM to select whether to think or not is unreliable, as its ranking utility remains well below that of consistently applying \thinkmode.
This indicates that \textbf{\textit{current LLMs cannot reliably assess instance difficulty or the marginal benefit of additional reasoning}}.

However, across all datasets, the \textsc{Ideal} mode can achieve markedly higher NDCG than either single mode while using far fewer tokens than \thinkmode.
For example, on MovieLens, NDCG increases by \textbf{32\%} while tokens decrease by \textbf{55\%}.
From the perspective of accuracy–efficiency trade-offs, the \textsc{Ideal} consistently surpasses both \nonthinkmode and \thinkmode baselines, emphasizing that \textbf{deciding \emph{when to reason} is as critical as \emph{how to reason}. }
Furthermore, Figure~\ref{fig:model_consistency} compares ideal routing decisions across different LLMs (Qwen3-4B/14B) on the same MS-MARCO test set.
The results reveal substantial cross-model variation: the same instance may be judged more suitable for \nonthinkmode by one LLM but for \thinkmode by another.
These observations motivate the design of a \textbf{\textit{model-aware, per-instance router}}, which selects between \nonthinkmode and \thinkmode \emph{before} inference to retain the benefits of \thinkmode when beneficial while avoiding unnecessary computation otherwise.

\begin{figure*}[t]
\setlength{\abovecaptionskip}{6pt}
\centerline{
\includegraphics[width=0.95\textwidth]{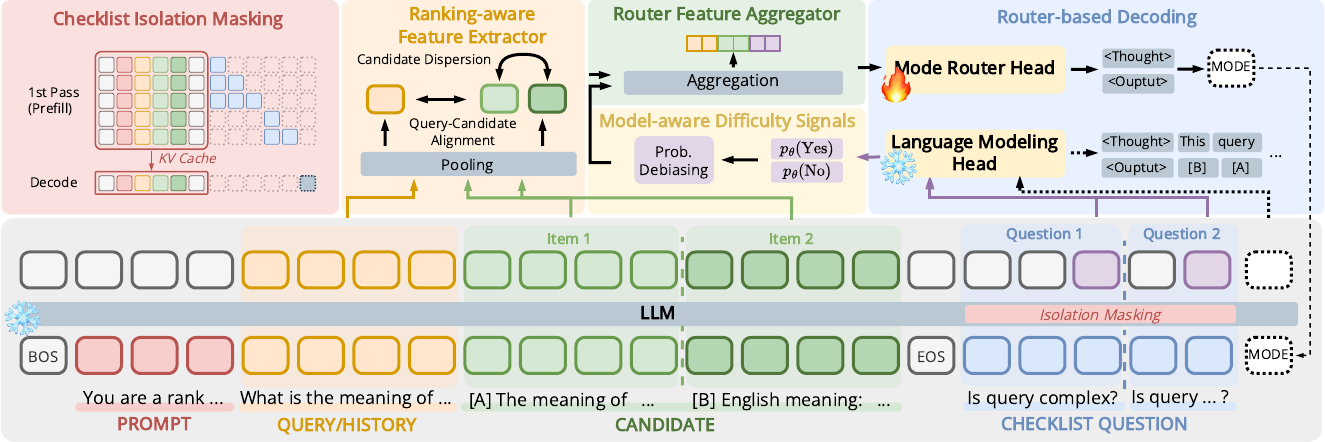}}
\caption{An overview of our lightweight, model-aware, per-instance router framework.}
\label{fig:framework}
\end{figure*}
\section{Method}
Building on the above motivation, we next address \textit{how to design} such a reasoning router.
Figure~\ref{fig:framework} provides an overview of our proposed framework, which comprises three key components.
First, to represent the structural complexity of each instance, we extract \textbf{ranking-aware features} from the LLM’s semantic embeddings that quantify contextual ambiguity and candidate dispersion.
Second, to model the LLM’s intrinsic perception of task difficulty, we employ a masked checklist of diagnostic yes/no questions whose response probabilities serve as \textbf{model-aware difficulty signals}, capturing the model’s internal confidence in processing the instance.
Finally, both feature types are concatenated and fed into a \textbf{lightweight router head}, which outputs a control token determining whether to apply \nonthinkmode or \thinkmode.
During deployment, the router leverages the \textbf{validation Pareto frontier} to flexibly select operating policies, allocating computation to instances expected to yield the greatest benefit from \thinkmode under system constraints.
All components execute within a \emph{single forward pass} of the backbone LLM, without modifying its architecture or decoding process.
This design preserves efficiency while adaptively allocating reasoning effort, achieving a balanced accuracy-efficiency trade-off.

\vspace{-0.2cm}
\subsection{Ranking-Aware Feature Extractor}
\hz{Prior studies have shown that ranking uncertainty is closely associated with both the intrinsic ambiguity of queries or user intents~\cite{yano2016quantifying, wang2025intent} and subtle relational dependencies among candidate items in determining their relevance ordering~\cite{li2025matching}. 
However, during inference, LLMs often struggle to directly assess the inherent complexity of a ranking instance from rich and unstructured textual descriptions.
To address this limitation, we construct a \textit{ranking-aware feature extractor} that quantifies the difficulty of different ranking instances. 
Motivated by the observation that ranking difficulty often stems from overlapping or weakly distinguishable representations, we first map the context and candidates into a shared semantic space using LLM embeddings, and then derive high-level statistical features to summarize their structural properties.}

\hz{Specifically, we extract segment embeddings for the \emph{context} (\(\mathcal{T}_C\)) and each \emph{candidate item \(i\)} (\(\mathcal{T}_i\)) from the final hidden-state layer \(S\in\mathbb{R}^{T\times d}\), thereby grounding feature computation in the model’s internal representational geometry.
Let \(\mathbf{S}\in\mathbb{R}^{T\times d}\) denote the last hidden states of the LLM over an input of \(T\) tokens. 
We define \(\mathcal{T}_C\subseteq\{1,\ldots,T\}\) and \(\mathcal{T}_i\subseteq\{1,\ldots,T\}\) as the \emph{index sets of tokens} that belong to the \emph{context} segment \(C\) and the \(i\)-th \emph{candidate} segment, respectively.
We then obtain segment embeddings by pooling over the corresponding rows of \(\mathbf{S}\):}
\begin{equation}
\mathbf{e}_C \;=\; \mathrm{pooling}\big(S_{\mathcal{T}_C}\big), 
\qquad
\mathbf{e}_i \;=\; \mathrm{pooling}\big(S_{\mathcal{T}_i}\big).
\vspace{-0.1cm}
\end{equation}
For IR task, \(C=Q\) and \(\mathbf{e}_C\) denotes the query embedding.
For Rec task, \(C=H=\{h_t\}_{t=1}^{m}\) and we additionally extract embedding \(\mathbf{e}_{h_t}=\mathrm{pooling}\big(S_{\mathcal{T}_{h_t}}\big)\) for each history-item.
\hz{These embeddings are used to derive high-level statistical features that quantify the structural properties of each ranking instance.}

\hz{Motivated by the intrinsic sources of difficulty in ranking tasks, we organize the feature set along two complementary dimensions:
\textbf{(i) Context/Candidate complexity.}
This dimension characterizes the internal variability of a ranking instance, encompassing both the coherence of the context (e.g., query specificity or stability of user interests) and the dispersion of the candidate set.
Features in this category measure the concentration or drift of context representations and the distributional spread of candidates in the embedding space, thereby reflecting the overall structural ambiguity of the ranking instance.
\textbf{(ii) Context–candidate alignment.}
This dimension assesses the degree of alignment between the context and candidate representations, as well as how decisively top-ranked items can be distinguished.
High alignment suggests clear relevance cues amenable to \nonthinkmode, whereas low alignment signals uncertainty in relevance estimation, under which explicit reasoning in \thinkmode may yield greater benefit~\cite{jiang2023llm}.}

\hz{To further select a compact yet robust feature subset, we adopt a two-stage refinement strategy that combines lightweight model-based feature selection with cross-setting consistency analysis. 
In the first stage, we train a lightweight predictor with sparsity-inducing regularization on the routing objective, enabling feature relevance to be assessed through learned coefficients or importance scores. 
This step filters out features with limited predictive contribution while retaining those most informative for estimating the expected utility of applying reasoning.
In the second stage, we examine the retained features across multiple datasets and backbone LLMs to assess their robustness. 
Only features that exhibit consistent predictive signals across different settings are preserved, whereas those showing unstable or dataset-specific effects are pruned. 
By combining predictive relevance with cross-setting consistency, we obtain a compact feature set\footnote{Due to space limitations, we provide the feature set in our open-source code.} that supports effective and robust routing, without introducing significant additional computational overhead.}

\subsection{Model-Aware Difficulty Signals}
While ranking-aware features characterize the intrinsic complexity of each instance, they do not capture model-specific competence or the subjective perception of difficulty across different LLMs (as illustrated in Figure~\ref{fig:selective-reasoning}).
Inspired by prior work~\cite{weirocketeval, pereira2024check}, we introduce a \textbf{checklist-based probing mechanism} that elicits the LLM’s internal perception of instance difficulty, thereby serving as a complementary signal to ranking-aware features.

\hz{We insert a small set of domain-tailored diagnostic questions into the prompt suffix as a masked checklist. 
Our core principle is to elicit the LLM’s responses without interfering with the ranking task, thereby providing model-aware signals for downstream routing. 
For instance, one question asks: \textit{``Is the query single-intent and unambiguous (no competing aspects)?''}, which captures the clarity of the input context. 
We then derive the model’s responses by extracting the \emph{Yes/No} probabilities from the hidden states at the \texttt{``Answer:''} anchor via the LM head, reflecting the model’s confidence in handling the given ranking instance.}

\hz{To construct the checklist, we adopt a systematic construction pipeline that balances coverage and compactness. 
We first leverage ChatGPT~\cite{achiam2023gpt} as a generative tool to produce an initial pool of candidate diagnostic questions, each designed to target a distinct aspect of ranking difficulty (e.g., contextual clarity, candidate ambiguity, or relevance separability).
The guiding principle at this stage is broad coverage, ensuring that candidate questions span diverse dimensions of potential difficulty.
From this initially over-complete set, we then apply a reflective elicitation strategy~\cite{madaan2023self, wang2025re2llm} to refine the checklist.
Specifically, by probing the LLM on representative ranking instances and analyzing the stability and redundancy of its responses, we identify questions that encode overlapping signals or exhibit limited discriminative value. 
Questions that provide complementary and diverse signals of model uncertainty are retained, whereas redundant or weakly informative ones are filtered out.}

\hz{However, prior work and our observations reveal that LLMs often exhibit a systematic \textit{yes-bias}, whereby semantically inverted questions (e.g., “Is this difficult?” vs.\ “Is this easy?”) can both elicit high \textit{yes} probabilities~\cite{zheng2023judging}.
Such directional bias undermines the reliability of checklist-derived signals and can unintentionally bias the model’s reasoning preference.
To mitigate this issue, we construct \textbf{paired, direction-balanced questions} to enhance the checklist, with one phrased to align with \nonthinkmode and the other with \thinkmode tendencies (e.g., single-intent vs.\ multi-intent queries).
By aggregating the paired responses, the router obtains more symmetric and robust evidence.\footnote{Due to space constraints, the complete checklist is provided in our open-source code.}}

\begin{tcolorbox}
\footnotesize
\textit{``Is the query single-intent and unambiguous (no competing aspects)?''}\\[2pt]
\textit{``Is the query ambiguous or multi-intent requiring aspect disambiguation?''}
\end{tcolorbox}

\hz{Building on the diagnostic question design, we integrate the checklist into the model input to extract model-aware difficulty signals without interfering with the ranking generation process.
A compact set of questions is appended to the prompt as a suffix and organized into independent blocks, one per question.
To prevent cross-interference, we apply a \textit{block-diagonal causal mask} that restricts each probe to attend only to the shared prefix and its own tokens, while blocking any interaction with other probes.
This masking enforces strict probe isolation, ensuring that checklist probing has no effect on ranking generation, while preserving efficiency via a single concatenated forward pass.}
Implementation details of checklist-based probing are provided in Appendix.

\hz{In summary, the checklist probing mechanism offers an efficient and non-intrusive way to obtain \textit{model-aware difficulty signals} by eliciting the LLM’s own uncertainty through isolated diagnostic questions. 
These signals complement the ranking-aware features and provide the router with additional evidence for determining when explicit reasoning is likely to be beneficial.}

\subsection{Reasoning Router Head}
\hz{With both ranking-aware features and checklist-derived model-aware difficulty signals available, we train a lightweight reasoning router \(g_{\phi}\) that selects between \nonthinkmode and \thinkmode on a per-instance basis. 
A key challenge in this setting is that the potential benefit of explicit reasoning is not binary but varies continuously across ranking instances. 
As shown in Figure~\ref{fig:pie_case}, the performance gap between \nonthinkmode and \thinkmode spans a wide range, from negligible or even negative impact to substantial gains. 
This suggests that \textit{the utility of explicit reasoning is inherently graded rather than categorical}.
Motivated by this observation, we formulate the routing objective as a regression problem that predicts a continuous utility margin in favor of \thinkmode over \nonthinkmode. 
By explicitly modeling the expected magnitude of reasoning benefit, the router can make finer-grained routing decisions that align more closely with the continuous nature of ranking difficulty and reasoning gains. 
This formulation also naturally supports flexible, cost-aware routing policies, in which reasoning is selectively allocated based on predicted utility rather than rigid mode assignments. 
We next describe the training procedure of the router \(g_{\phi}\) in detail.}


\textbf{Labeling Objective (Advantage of \thinkmode over \nonthinkmode).}
We train the router using a \textbf{compute-aware advantage} objective that quantifies the benefit of employing \thinkmode over \nonthinkmode for each instance \(p\):
\begin{equation}
    A_p \;=\; \big(U_p^{\thinkmode} - U_p^{\textsc{Non}}\big)\;-\;\lambda \big(T_p^{\thinkmode} - T_p^{\textsc{Non}}\big),
\end{equation}
where \(U\) denotes the ranking utility (e.g., NDCG), \(T\) represents the inference cost (e.g., tokens), and \(\lambda>0\) controls the trade-off between accuracy and efficiency.
This formulation encourages the router to favor \thinkmode only when its expected performance improvement justifies the additional computational cost.

\textbf{Feature Selection.}
After feature extraction, each instance is represented by a feature vector \(\mathbf{x}_p\) that integrates both \emph{ranking-aware} statistics and \emph{model-aware difficulty signals}.
However, the resulting feature space is high-dimensional and may contain redundant or confounding variables.
To enhance generalization and interpretability, we perform a feature-selection stage \emph{prior} to training the router, aiming to reduce dimensionality and suppress noisy or irrelevant signals.
Inspired by \textit{ManiFeSt}~\cite{cohen2023few}, we analyze feature dependencies within each class, i.e., among instances with similar labels \(A_p\), rather than relying solely on feature-label correlations. 
This intra-class relational analysis yields a compact yet informative subset that minimizes redundancy and improves the stability and robustness of the router \(g_{\phi}\) across inputs and model scales.


\textbf{Router Head Training.}
To train the router head, we jointly consider predictive accuracy and structural consistency with the compute-aware objective.
Given feature vectors \(\mathbf{x}\), 
we train a router head \(g_\phi:\mathbf{x}\mapsto A\) (implemented as a gradient-boosted regressor~\cite{mohan2011web}) to estimate the expected advantage of \thinkmode over \nonthinkmode.
To align learning with the efficiency objective, we impose a \textbf{monotonicity constraint} on the extra-cost feature \(\Delta T_p = T_p^{\thinkmode}-T_p^{\textsc{Non}}\), ensuring that the predicted advantage does not increase with higher computational cost.
Let \(N_p\) denote the total number of instances.
The model parameters \(\phi\) are optimized under this structural constraint using a weighted squared loss:
\begin{equation}
\min_{\phi}\;\; \frac{1}{N_p}\sum_{p=1}^{N_p} w_p\,\big(f_\phi(\mathbf{x}_p)- A_p\big)^2
\quad
\text{s.t.}\quad
\frac{\partial f_\phi}{\partial \Delta T} \le 0,
\end{equation}
where \(w_p\) denotes robustness weights against noisy instances.

\vspace{-0.2cm}
\subsection{Policy Selection on the Pareto Frontier.}
\hz{After training the router, a practical question remains: \emph{how to select an operating policy that best balances accuracy and computational cost in real-world deployments?}
Different applications impose different constraints—some prioritize low latency or strict token budgets, while others require high-quality rankings even at higher computational cost.
Therefore, a single fixed threshold is insufficient; instead, we require a principled mechanism that exposes the full accuracy–efficiency trade-off and enables practitioners to choose the configuration best aligned with their operational requirements.}

\hz{To support this selection process, we employ the \textbf{validation Pareto frontier}~\cite{shah2016pareto}, which offers a characterization of the trade-offs between token cost \(T\) and ranking utility \(U\). 
The frontier consists of the non-dominated operating points \((T_j, U_j)\), where utility cannot be improved without incurring additional computational cost.
As shown in Figure~\ref{fig:pareto_curve}, this representation offers an interpretable landscape of optimal accuracy–efficiency regimes: points located toward the right indicate lower-cost, faster inference settings, whereas points near the top correspond to higher-utility. 
Analyzing the frontier enables practitioners to quantify marginal cost–utility trade-offs, such as the token expenditure required to obtain utility, and to identify operating points that maximize utility–cost efficiency under specific deployment constraints.}

\hz{To translate high-level operational goals into actionable routing behavior, we define four representative \emph{deployment anchors} and describe their corresponding optimization criteria.
Each anchor specifies a target cost (or utility) level, from which we determine the density parameter \(\eta\) used in the routing rule.}

\medskip
$\blacklozenge$ \textbf{\textit{Policy I: Knee (efficiency-first).}} 
\hz{This policy is intended for applications in which \emph{latency and token budget} are primary operational constraints. 
Its objective is to cease allocating additional computation once utility gains exhibit diminishing returns. 
We first normalize cost and utility:}
\begin{equation}
\tilde T_j = \frac{T_j - T_{\min}}{T_{\max} - T_{\min}}, \qquad
\tilde U_j = \frac{U_j - U_{\min}}{U_{\max} - U_{\min}},
\end{equation}
and select the point farthest from the anti-diagonal:
\begin{equation}
j^\star = \arg\max_j d_j^\perp, 
\qquad
d_j^\perp = \frac{|\tilde T_j + \tilde U_j - 1|}{\sqrt{2}}.
\end{equation}
\hz{This identifies the ``knee", the point at which further computation provides diminishing returns.
We then solve for \(\eta_{\mathrm{knee}}\) such that the average token cost matches the implied budget \(\bar T_{\mathrm{knee}} = T_{j^\star}\):}
\begin{equation}
\min_{\eta\ge0} \big(\mathbb{E}{\mathrm{val}}[T(a(\eta))] - \bar T_{\mathrm{knee}}\big)^2.
\end{equation}
\hz{This policy is ideal when cost savings matter more than marginal accuracy improvements.}

\medskip
$\blacklozenge$ \textbf{\textit{Policy II: Utopia (ideal balance).}} 
\hz{Many ranking applications (e.g., search re-ranking, personalized feeds) require a \emph{balanced} trade-off that achieves strong utility at a reasonable computational cost.
This policy identifies the point closest to the ideal of low cost and high utility:}
\begin{equation}
j^\star = 
\arg\min_j \left\{
w_T \tilde T_j^2 + w_U (1 - \tilde U_j)^2
\right\},
\end{equation}
\hz{Weights \(w_T, w_U\) allow tuning toward either efficiency or accuracy.
We then compute \(\eta_{\mathrm{u_t}}\) to reproduce the implied token level.
This policy suits general-purpose ranking systems deployed at scale.}

\medskip
$\blacklozenge$ \textbf{\textit{Policy III: Epsilon (utility-constrained).}} 
\hz{Some applications require meeting a strict utility threshold, such as fairness-sensitive re-rankers.
In this setting, the objective is to \emph{meet a specified utility target while minimizing computational cost}.
Given a utility target:}
\begin{equation}
U_{\mathrm{target}} = U_{\mathrm{base}} + \epsilon,
\end{equation}
we select the lowest-cost frontier point that satisfies \(U_j \ge U_{\mathrm{target}}\) and then solve:
\begin{equation}
\min_{\eta \ge 0} \mathbb{E}_{\mathrm{val}}[T(a(\eta))]
\quad
\text{s.t.}\quad
\mathbb{E}_{\mathrm{val}}[U(a(\eta))] \ge U_{\mathrm{target}}.
\end{equation}
\hz{This policy is well suited to scenarios in which a minimum acceptable utility level must be guaranteed.}

\medskip
$\blacklozenge$ \textbf{\textit{Policy IV: UMax (utility-first).}}  
\hz{In offline evaluation, user-modeling tasks, or premium ranking services, maximizing utility is the primary objective, even at increased computational cost.
Accordingly, this policy selects:}
\begin{equation}
j^\star = \arg\max_j U_j, \qquad \bar T_{\mathrm{UMax}} = T_{j^\star}.
\end{equation}
\hz{We then find \(\eta_{\mathrm{UMax}}\) that achieves the matching token level.
This corresponds to a low-\(\eta\) regime that sends most instances to \thinkmode.}

\medskip
\textbf{Routing rule.}
\hz{All policies ultimately determine a density parameter \(\eta\) used in the cost-aware routing rule:}
\begin{equation}
\label{eq:router_rule}
a_p(\eta)=
\begin{cases}
\thinkmode, & A_p - \eta\, (T_p^{\thinkmode} - T_p^{\textsc{Non}}) \ge 0,\\
\nonthinkmode, & \text{otherwise},
\end{cases}
\end{equation}
\hz{which compares the predicted calibrated advantage \(A_p\) against the cost penalty for each ranking instance \(p\).  
The selected \(\eta\) is frozen for deployment, ensuring stable and auditable behavior aligned with real-world latency and budget constraints.}

\section{Experiment}
\label{sec:exp}

\begin{table*}[t]
\footnotesize
\setlength{\abovecaptionskip}{7pt}
\setlength\tabcolsep{3pt}
\setul{1pt}{0.4pt}
\setlength{\cmidrulekern}{.2em}
\renewcommand{\arraystretch}{0.9}\
\centering
\caption{
\centering
Ranking utility and token cost across datasets and LLM backbones.
All fractional metrics \\ are scaled by a factor of 100 for readability and reported as the average over five independent runs.
}
\label{tab:average_results}
\begin{tabular}{c c c c c c c c c c c c c c c c c}
\toprule
\multicolumn{2}{c}{} & \multicolumn{5}{c}{\textbf{MovieLens}} & \multicolumn{5}{c}{\textbf{Amazon-VG}} & \multicolumn{5}{c}{\textbf{MS-MARCO}} \\ 
\cmidrule(l){3-17} 
\multicolumn{2}{c}{\textbf{Model}} & R@5 & N@5 & R@10 & N@10 & Token & R@5 & N@5 & R@10 & N@10 & Token & Top1 & PwAcc & N@5 & N@10 & Token \\ 
\midrule
\multicolumn{2}{c}{\textbf{ARM-3B}} & 8.36 & 5.06 & 17.31 & 7.91 & 469 & 9.94 & 6.17 & 18.75 & 8.97 & 513 & 15.50 & 56.31 & 72.36 & 78.45 & 716 \\ 
\multicolumn{2}{c}{\textbf{ARM-7B}} & 13.75 & 8.91 & 22.18 & 11.60 & 1054 & 9.51 & 6.41 & 16.31 & 8.58 & 939 & 24.99 & 40.86 & 49.84 & 51.01 & 487 \\ 
\multicolumn{2}{c}{\textbf{ARM-14B}} & 19.33 & 13.36 & 29.87 & 16.74 & 1071 & 14.02 & 9.86 & 23.59 & 12.89 & 1090 & 40.46 & 62.50 & 77.40 & 79.03 & 1275 \\ 
\midrule
\multirow{5}{*}{\textbf{Qwen3-4B}} & \nonthinkmode & 20.44 & 14.76 & 32.29 & 18.59 & 48 & 19.09 & 14.44 & 27.12 & 17.01 & 63 & 46.48 & 70.27 & 85.62 & 87.74 & 55 \\
 & \thinkmode & {21.65} & 14.81 & {34.31} & {19.02} & 384 & 24.20 & 17.47 & 32.87 & 20.34 & 404 & {48.61} & {72.04} & {86.89} & {87.99} & 263 \\
\mydashline
 & \textsc{Random} & 21.35 & {15.02} & 33.34 & 18.88 & 157 & {21.36} & {15.45} & {30.99} & {18.47} & 303 & 47.70 & {71.13} & 86.21 & 87.88 & 163 \\
 & \textsc{Self-Select} & 19.93 & 13.91 & 31.52 & 17.64 & 64 & 19.15 & 14.49 & 27.16 & 17.04 & 125 & 45.45 & 70.52 & 85.99 & 87.90 & 160 \\
 & Ours & \textbf{22.79} & \textbf{16.22} & \textbf{35.70} & \textbf{20.22} & 194 & \textbf{24.90} & \textbf{18.08} & \textbf{34.06} & \textbf{21.10} & 321 & \textbf{48.73} & \textbf{72.11} & \textbf{87.02} & \textbf{88.26} & 65 \\ 
\midrule
 \multirow{5}{*}{\textbf{Qwen3-8B}} & \nonthinkmode  & 22.74 & 15.78 & 34.86 & 19.74 & 65 & 18.83 & 14.25 & 26.61 & 16.90 & 71 & 48.64 & 71.40 & 86.04 & {88.31} & 55 \\
 & \thinkmode & 23.67 & 16.21 & 35.52 & 20.15 & 292 & {23.54} & {17.53} & {32.71} & {20.58} & 285 & {49.34} & {72.74} & {86.83} & 87.71 & 417 \\
 \mydashline
 & \textsc{Random} & 24.36 & 16.40 & 36.45 & 20.66 & 196 & 22.09 & 16.65 & 30.07 & 19.40 & 173 & 48.75 & 72.04 & 86.36 & 87.78 & 183 \\
 & \textsc{Self-Select} & {24.39} & {16.69} & {37.32} & {20.90} & 198 & 20.53 & 15.79 & 28.06 & 18.37 & 66 & 47.28 & 71.42 & 86.36 & 88.49 & 60 \\
 & Ours & \textbf{25.32} & \textbf{17.34} & \textbf{38.02} & \textbf{21.65} & 175 & \textbf{24.28} & \textbf{18.03} & \textbf{33.07} & \textbf{20.98} & 242 & \textbf{49.53} & \textbf{73.05} & \textbf{87.45} & \textbf{88.68} & 68 \\ 
\midrule
 \multirow{5}{*}{\textbf{Qwen3-14B}} & \nonthinkmode  & 23.16 & 15.86 & 35.15 & 19.70 & 56 & 23.76 & 18.20 & 33.37 & 21.27 & 63 & 48.11 & 71.31 & 83.76 & 85.56 & 68 \\
 & \thinkmode & 24.78 & 17.03 & 36.33 & 20.75 & 368 & {29.36} & {21.16} & {41.14} & {24.96} & 563 & {52.85} & {75.32} & {88.45} & {90.94} & 319 \\
 \mydashline
 & \textsc{Random} & {25.02} & {16.97} & {38.14} & {21.18} & 296 & 28.60 & 20.89 & 40.00 & 24.55 & 303 & 50.75 & 73.72 & 86.52 & 88.36 & 189 \\
 & \textsc{Self-Select} & 22.67 & 15.32 & 35.58 & 19.47 & 58 & 26.27 & 19.29 & 36.95 & 22.71 & 190 & 39.94 & 58.32 & 68.05 & 69.59 & 249 \\
 & Ours & \textbf{25.68} & \textbf{18.61} & \textbf{39.26} & \textbf{21.63} & 254 & \textbf{30.32} & \textbf{21.99} & \textbf{41.14} & \textbf{25.52} & 279 & \textbf{53.43} & \textbf{76.05} & \textbf{89.31} & \textbf{91.22} & 265 \\ 
 \midrule
\multirow{5}{*}{\textbf{Gemma3-12B}} & \nonthinkmode & 29.86 & 21.20 & 44.74 & 25.96 & \textbf{63} & 29.66 & 21.04 & 42.02 & 24.99 & \textbf{69} & 36.96 & 63.84 & 74.89 & 77.32 & \textbf{54} \\
 & \thinkmode & 25.12 & 17.09 & 37.96 & 21.22 & 234 & 30.78 & 21.87 & 43.62 & 26.00 & 210 & \textbf{46.49} & \textbf{74.58} & \textbf{87.88} & \textbf{90.26} & 253 \\
  \mydashline{}
 & \textsc{Random} & 26.81 & 18.67 & 40.65 & 23.12 & 144 & 29.51 & 20.89 & 41.98 & 24.42 & 139 & 41.60 & 68.89 & 80.96 & 83.37 & 146 \\
 & \textsc{Self-Select} & 29.34 & 20.36 & 43.79 & 24.99 & 97 & 29.91 & 21.18 & 42.33 & 25.14 & 88 & 41.86 & 69.30 & 81.58 & 83.99 & 152 \\
 & Ours & \textbf{30.25} & \textbf{21.36} & \textbf{44.81} & \textbf{26.12} & 83 & \textbf{32.20} & \textbf{22.91} & \textbf{45.53} & \textbf{27.13} & 150 & 45.98 & 74.11 & 87.32 & 89.69 & 167 \\ 
 \midrule
 \multirow{5}{*}{\textbf{GPT-Oss-20B}} & \nonthinkmode & 26.72 & 19.72 & 35.34 & 23.80 & \textbf{43} & 28.89 & 22.21 & 36.56 & 26.69 & \textbf{43} & 45.24 & 62.03 & 73.75 & 75.43 & \textbf{70} \\
 & \thinkmode & 30.08 & 20.82 & 42.31 & 25.01 & 360 & 33.70 & 24.20 & 44.45 & 28.17 & 388 & 51.14 & 72.41 & 86.93 & 88.43 & 234 \\
 \mydashline
 & \textsc{Random} & 28.80 & 20.48 & 38.72 & 24.46 & 237 & 30.20 & 22.64 & 40.42 & 26.95 & 261 & 48.17 & 66.88 & 79.92 & 81.52 & 151 \\
 & \textsc{Self-Select} & 26.82 & 19.75 & 35.83 & 24.15 & 46 & 28.65 & 22.19 & 36.81 & 27.01 & 46 & 46.16 & 62.95 & 74.69 & 76.37 & 72 \\
 & Ours & \textbf{30.15} & \textbf{21.03} & \textbf{42.40} & \textbf{25.25} & 272 & \textbf{33.91} & \textbf{24.52} & \textbf{44.69} & \textbf{28.91} & 282 & \textbf{51.25} & \textbf{72.47} & \textbf{86.94} & \textbf{88.47} & 182 \\ 
 \bottomrule
\end{tabular}
\end{table*}

In this section, we present extensive experiments on three public datasets and five open-source LLM backbones with varying parameter scales.
Our goal is to demonstrate that our proposed lightweight, model-aware router yields better ranking utility and lower inference cost than always using either \nonthinkmode or \thinkmode. 
We organize this section around three research questions.

\textbf{\textit{RQ1:}} \textit{Does the proposed router improve ranking utility while reducing token usage compared with baselines?}

\textbf{\textit{RQ2:}} \textit{How does a Pareto-frontier–based policy achieve better accuracy–efficiency trade-offs than single-mode baselines?}

\textbf{\textit{RQ3:}} \textit{How do our two components shape the router’s decisions across datasets in ablation experiments?}

\textbf{Datasets.}
For the IR task, we use \textbf{MS-MARCO}~\cite{pradeep2023rankzephyr}, which provides queries with variable-length candidate pools. 
And evaluation is conducted using the original pools without augmentation.
For the Rec task, we evaluate on \textbf{MovieLens} (movies)~\cite{harper2015movielens} and \textbf{Amazon-VG} (video games)~\cite{ni2019justifying}, both containing user–item interaction logs.
For each user, interactions are sorted by \textit{day}, and a leave-one-out protocol is applied: the last item is held out as the target, and previous interactions form the history~\cite{zhang2025hierarchical}.
We uniformly sample 49 negatives from non-interacted items and pair them with the target to construct a fixed-size candidate set of 50 per instance, following the setting of~\cite{wang2025re2llm}.
For each dataset, we split all data into train, validation, and test sets at a 6:2:2 ratio.


\textbf{Backbone Models.} We evaluate five open-source LLM backbones spanning a broad range of model scales: \textbf{Qwen3-4B}, \textbf{Qwen3-8B}, \textbf{Qwen3-14B}~\cite{qwen3technicalreport}, \textbf{Gemma3-12B}~\cite{gemma_2025}, and \textbf{GPT-Oss-20B}~\cite{openai2025gptoss120bgptoss20bmodel}.
We use their instruction-tuned chat variants without any parameter updates, and apply \emph{exactly the same prompt} across reasoning modes (Figure~\ref{fig:prompt_all_2}); all runs share the same decoding hyperparameters.

\textbf{Selected Baselines.}
We use the ranking results of the backbone model under four inference modes—\nonthinkmode, \thinkmode, \textsc{Random}, and \textsc{Self-select} as baselines for comparison. 
In addition, we include comparisons with the \textbf{ARM} family of open-source LLMs, which represent state-of-the-art (SOTA) work in adaptive reasoning for general task settings~\cite{wu2025arm}, to highlight the superior performance of our proposed framework.
There also exist studies~\cite{wan2025adapthink, aggarwal2025l1} that enhance adaptive reasoning in mathematical and coding tasks by fine-tuning LLMs on carefully constructed datasets.
However, due to the lack of publicly available datasets for ranking tasks and the absence of open-sourced fine-tuned models, direct comparison with these approaches is not feasible.

\textbf{Evaluation and Metrics}.
We evaluate ranking performance using four metrics. 
\textit{Top-1 Agreement}~\cite{ekstrom2019sequential} \textit{(Top1)} checks whether the top predicted item matches the ground-truth. 
\textit{Pairwise Accuracy} \textit{(PwAcc)}~\cite{beutel2019fairness}  measures order consistency between prediction and ground truth over common items. 
\textit{Recall@k (R@k)}~\cite{wang2025enhancing} quantifies coverage by the proportion of ground-truth items retrieved within the top-k.
\textit{NDCG@k (N@k)}~\cite{jarvelin2002cumulated} captures position-sensitive ranking quality, rewarding correct items appearing earlier. 

\textbf{Implementation Details.}
All experiments were conducted on a Linux server with $4\times$ NVIDIA RTX 6000 Ada Generation GPUs, each providing 46GB of memory. The GPU driver version was 570.124.04.

\subsection{Experimental Results and Analysis}
\subsubsection{Answer to RQ1:}
In Table~\ref{tab:average_results}, we report results for the \textit{UMax} policy of all models across three datasets. 
For the Rec task, we employ \textit{Recall} and \textit{NDCG} to evaluate recommendation accuracy. 
For the IR task, we use \textit{Top1}, \textit{PwAcc}, and \textit{NDCG} to assess ranking utility.
Higher values indicate better results. 
In all experiments, we report the number of inference tokens as a measure of inference efficiency, with lower tokens reflecting higher efficiency. 
Bolded values in the table denote the best results under the same backbone. 

\emph{\textbf{First, our router attains higher ranking utility than \textit{\thinkmode} while consuming substantially fewer tokens in most settings.}}
Across datasets and model scales, our router delivers a clear accuracy–efficiency advantage. 
For example, on MovieLens with Qwen3-4B, N@10 increases from \textit{0.190} to \textit{0.202} (\textbf{+6.3\%}), while the average token count drops by \textbf{49.5\%}. 
A similar pattern holds for Amazon-VG (\textit{0.203}→\textit{0.211}, \textbf{+3.9\%} N@10, \textbf{–20.5\%} tokens) and MS-MARCO (\textit{0.880}→\textit{0.882}, \textbf{–75.3\%} tokens).  
However, on Gemma3-12B, where \thinkmode substantially outperforms \nonthinkmode, our router does not surpass \thinkmode in ranking utility; nevertheless, it matches \thinkmode to within \textbf{99.37\%} while reducing token usage by \textbf{34\%}.
These results indicate that the \textit{model-aware router} preserves the benefits of reasoning while avoiding redundant computation, achieving the best utility-cost trade-off overall across our settings.

\begin{figure*}[t]
  \setlength{\abovecaptionskip}{2pt}
  \centering
  \subfigure[MovieLens]{
    \includegraphics[width=0.3\linewidth]{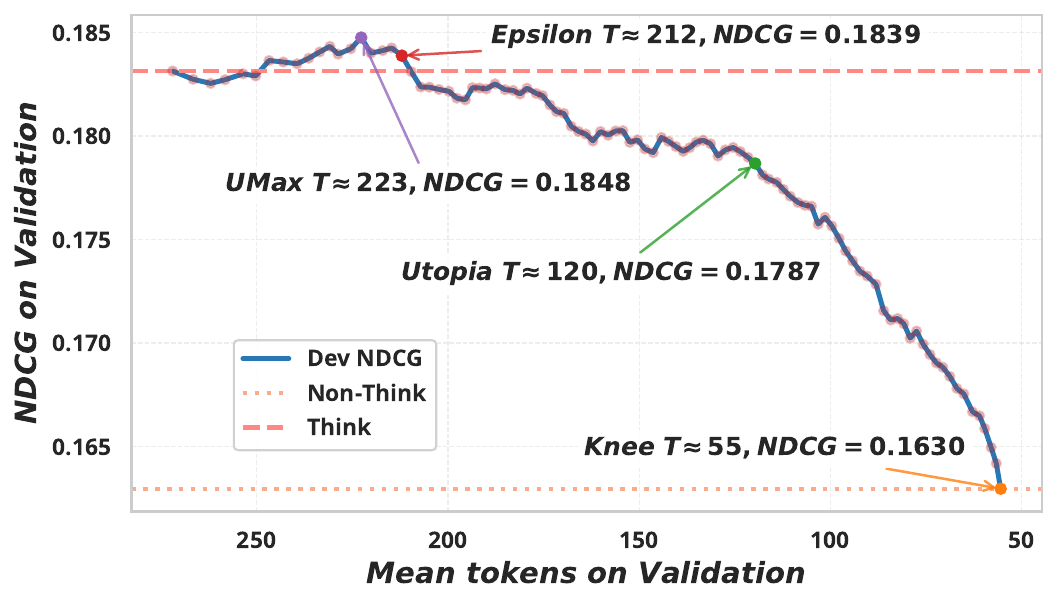}
    \label{fig:budget_ml-1m}
  }\hfill
  \subfigure[Amazon-VG]{
  \includegraphics[width=0.3\linewidth]{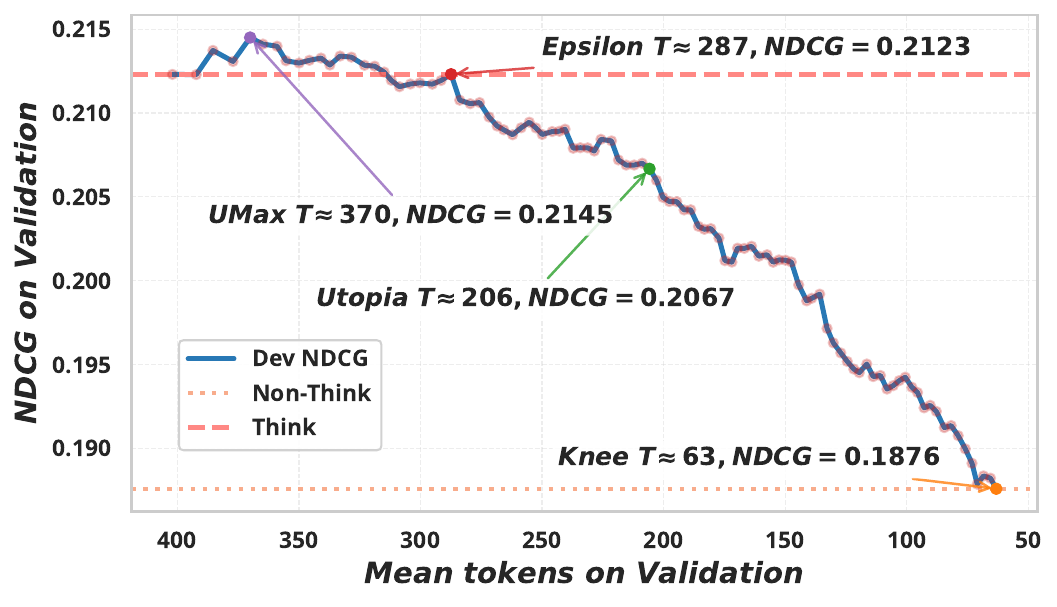}
    \label{fig:budget_amazon-game}
  }\hfill
  \subfigure[MS-MARCO]{
  \includegraphics[width=0.3\linewidth]{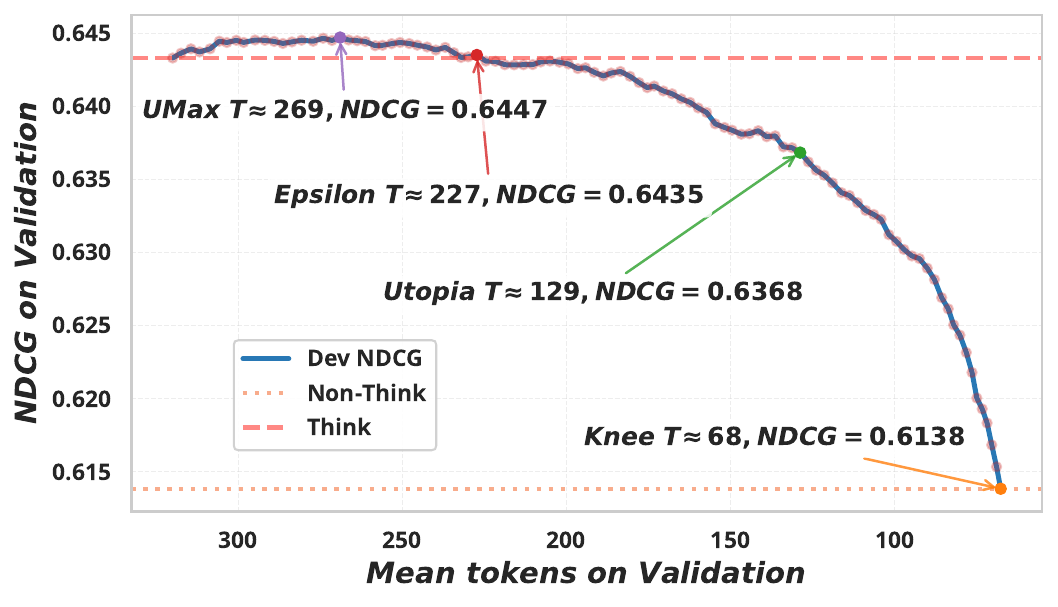}
    \label{fig:budget_rank_ir}
  }
  \caption{Pareto frontier curves on the validation sets of three ranking datasets, showing the trade-off between ranking utility (NDCG@10) and token consumption. Always-\nonthinkmode and Always-\thinkmode serve as baselines. Key operating points—including Knee, Utopia, Epsilon, and UMax—are highlighted.}
  \label{fig:pareto_curve}
\end{figure*}

\emph{\textbf{Second, while larger backbones raise the overall utility ceiling, per-instance routing remains essential and generalizes across model families.}}
As model size increases, both \nonthinkmode and \thinkmode generally improve (e.g., MovieLens N@10 for \thinkmode: \textit{0.190} $\to$ \textit{0.202} $\to$ \textit{0.207} from Qwen3-4B to 14B). 
However, the phenomenon identified in Section~\ref{sec:motivation} persists: \nonthinkmode still wins on non-trivial subsets. 
For instance, on MS-MARCO with Qwen3-8B, \nonthinkmode slightly surpasses \thinkmode (\textit{0.883} vs.~\textit{0.877}) despite \thinkmode using $\sim$8 times more tokens (\textit{417} vs.~\textit{55}). 
Thus, a uniform \thinkmode mode remains inefficient even for larger LLMs. 
Our model-aware router preserves \thinkmode's gains when they are most beneficial while avoiding unnecessary cost, attaining consistent improvements across Qwen3-4B/8B/14B and generalizing to architecturally different backbones, including Gemma and GPT-style models.

\textbf{\emph{Third, our proposed method advances the accuracy-efficiency frontier across datasets and backbones.}}
Compared to heuristic baselines such as \textsc{Random} and \textsc{Self-select}, our proposed router attains dominant ranking utility under markedly lower token cost.
For example, on MovieLens with Qwen3-8B, N@10 rises from \textit{0.207} (\textsc{Random}) and \textit{0.209} (\textsc{Self-select}) to \textit{0.216} (Ours).
Moreover, when compared with the \textbf{ARM} family of adaptive reasoning models, we observe that ARM tends to \textit{classify nearly all instances as ``hard'' tasks, defaulting to long reasoning chains. }
They substantially increase inference cost while simultaneously degrading ranking utility, indicating that \textbf{such designs tailored for general tasks are ill-suited for ranking scenarios}. 
This further underscores the necessity of a model-aware reasoning router tailored for the ranking task as realized in our framework.

\begin{figure*}[ht]
  \centering
  \includegraphics[width=0.98\linewidth]{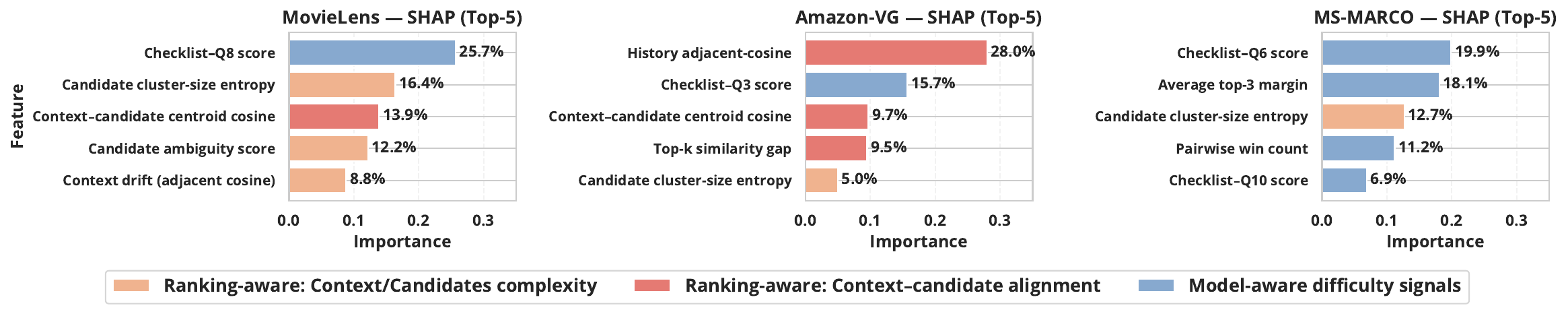}
  \vspace{-0.2cm}
  \caption{Top-5 feature importances of the mode router on the validation sets for three datasets (MovieLens, Amazon-VG, MS-MARCO). Bars are colored by feature family: \emph{ranking-aware features} in red and \emph{model-aware difficulty signals} in blue. Longer bars indicate stronger predictive power for the \thinkmode-vs-\nonthinkmode advantage.}
  \label{fig:feature-importance-multi}
  \vspace{-0.2cm}
\end{figure*}

\subsubsection{Answer to RQ2:}
Figure~\ref{fig:pareto_curve} shows the validation Pareto frontier, which demonstrates the flexibility of our router under deployment scenarios with varying degrees of cost sensitivity.
The x-axis denotes the mean tokens per instance and the y-axis indicates ranking utility (\textit{NDCG@10}).
The \emph{solid colored curve} represents the Pareto frontier obtained from a parameter sweep, while \emph{semi-transparent dots} mark all explored operating points.
Dotted and dashed horizontal lines denote \emph{\nonthinkmode} and \emph{\thinkmode} baselines, respectively.
Four labeled anchors summarize typical deployment regimes: \emph{Knee},
\emph{Utopia},
\emph{Epsilon},
and \emph{UMax}. 
The curve is constructed by computing per-instance advantages $A_p$ and applying the cost-aware routing rule in Equation~\ref{eq:router_rule}; sweeping the control parameter $\eta$ yields pairs $\big(T(\eta),U(\eta)\big)$.
Across three ranking datasets, the validation Pareto curves demonstrate that our proposed \emph{reasoning routing} framework consistently matches or surpasses single-mode baselines.
By using \thinkmode only when it is predicted to be beneficial, the router attains \thinkmode-level utility under substantially lower token budgets and, in several regions, even achieves higher utility.
The four annotated anchors correspond cleanly to distinct deployment regimes. 
The \textbf{Epsilon} (quality-constrained) point reaches the \thinkmode NDCG with \(\approx 10\text{--}30\%\) fewer tokens on MovieLens and Amazon-VG, and attains near-\thinkmode quality at low cost on MS-MARCO. 
The \textbf{Utopia} (balanced) point offers a strong default configuration, typically halving token usage relative to \thinkmode with only a small utility drop (e.g., \textit{0.1787} vs. \textit{0.183}). 
The \textbf{Knee} (efficiency-first) point lies just before diminishing returns, achieving \textbf{60–80\% savings} with controlled accuracy loss, making it suitable for latency-sensitive systems. 
Conversely, the \textbf{UMax} (accuracy-first) point often exceeds the peak NDCG of \thinkmode by avoiding harmful over-reasoning. 
Notably, we find that \nonthinkmode outperforms \thinkmode using Gemma3-12B with MovieLens, but our designed router remains effective when \thinkmode is not the better mode. 

\subsubsection{Answer to RQ3:}

As shown in Figure~\ref{fig:feature-importance-multi}, our SHAP-based feature-importance analysis across datasets reveals that \textit{effective reasoning routing is driven by task-specific cues}, and the router adapts its decision basis accordingly. 
On the two Rec datasets, all feature categories contribute. However, ranking-aware signals consistently rank as the most influential factors. 
These signals capture user context and candidate-set complexity (e.g., candidate cluster-size entropy) as well as context–candidate alignment (e.g., context-candidate centroid cosine similarity).
This pattern suggests that personalization and matching structure are central to deciding when \thinkmode is beneficial: \textbf{\textit{the router leverages stable user preferences and fine-grained context–item alignment to trigger \thinkmode primarily when additional reasoning is likely to alter the top-ranked items}}. 
In contrast, on the IR dataset, salient contributors shift toward model-aware difficulty signals derived from the checklist. 
These signals more directly reflect the LLM’s perceived query hardness and the expected marginal benefit of deeper reasoning. 
This shift is consistent with IR, where per-query complexity, rather than long-term preference, primarily determines the utility of \thinkmode.

To further quantify the contribution of these feature types, we conduct ablation studies by removing either the ranking-aware features (w/o ranking-aware) or the model-aware difficulty signals (w/o model-aware) and compare against the full router. 
We evaluate three relatively large LLM backbones on a Rec dataset and an IR dataset to improve robustness, with results summarized in Table~\ref{tab:ablation}.
Due to space constraints, we report a subset of metrics and observe consistent trends for the remaining metrics. 
To jointly capture effectiveness and efficiency, we define \(\textit{Trade-off} =\textit{N@10} -10^{-4}\times\textit{Token}\).
Overall, \textit{\textbf{removing either component consistently degrades performance}}, as reflected by lower ranking accuracy and a reduced Trade-off. 
This result indicates that both feature groups are essential for routing. 
Notably, the relative impact is task-dependent. 
On Rec task, removing ranking-aware features causes larger drops, suggesting that the router relies on preference and ranking-structure cues to identify ambiguous cases and allocate additional reasoning. 
In contrast, on IR task, model-aware difficulty signals contribute more, because queries vary substantially in ambiguity and information needs.
These difficulty probes help the router detect challenging instances and disambiguate borderline candidates, enabling it to spend computation where deeper reasoning yields the greatest benefit while conserving tokens on easier queries.

\begin{table}[h]
\footnotesize
\setlength{\abovecaptionskip}{3pt}
\setlength\tabcolsep{4pt}
\renewcommand{\arraystretch}{0.9}
\centering
\caption{Ablation study of feature groups for the router.}
\begin{tabular}{cccccc}
\toprule
\multicolumn{6}{c}{\textbf{Amazon-VG}} \\
\midrule
\textbf{Model} & \textbf{Variant} & \textit{R@10} & \textit{N@10} & \textit{Token} & Trade-off \\
\midrule
\multirow{3}{*}{\textbf{Qwen3-14B}}
& \textit{(w/o) ranking-aware} & 39.24 & 23.57 & 254 & 21.03 \\
& \textit{(w/o) model-aware} & \underline{40.58} & \underline{24.61} & 286 & \underline{21.75} \\
& \textit{Ours} & \textbf{41.14} & \textbf{25.52} & 279 & \textbf{22.73} \\
\midrule
\multirow{3}{*}{\textbf{Gemma3-12B}}
& \textit{(w/o) ranking-aware} & 43.95 & 26.04 & 117 & 24.87 \\
& \textit{(w/o) model-aware} & \underline{44.87} & \underline{26.91} & 144 & \underline{25.47} \\
& \textit{Ours} & \textbf{45.53} & \textbf{27.13} & 150 & \textbf{25.63} \\

\midrule
\multirow{3}{*}{\textbf{GPT-Oss-20B}}
& \textit{(w/o) ranking-aware} & 39.58 & 27.32 & 152 & 25.80 \\
& \textit{(w/o) model-aware} & \underline{39.83} & \underline{27.89} & 204 & \underline{25.85} \\
& \textit{Ours} & \textbf{44.69} & \textbf{28.91} & 282 & \textbf{26.09} \\
\midrule
\multicolumn{6}{c}{\textbf{MS-MARCO}} \\
\midrule
\textbf{Model} & \textbf{Variant} & \textit{PwAcc} & \textit{N@10} & \textit{Token} & Trade-off \\
\midrule
\multirow{3}{*}{\textbf{Qwen3-14B}}
& \textit{(w/o) ranking-aware} & \underline{75.88} & \underline{90.87} & 243 & \underline{88.44} \\
& \textit{(w/o) model-aware} & 75.45 & 90.44 & 236 & 88.08 \\
& \textit{Ours} & \textbf{76.05} & \textbf{91.22} & 265 & \textbf{88.57} \\

\midrule
\multirow{3}{*}{\textbf{Gemma3-12B}}
& \textit{(w/o) ranking-aware} & \underline{72.33} & \underline{87.51} & 152 & \underline{85.99} \\
& \textit{(w/o) model-aware} & 67.69 & 81.97 & 104 & 80.93 \\
& \textit{Ours} & \textbf{74.11} & \textbf{89.69} & 167 & \textbf{88.02} \\

\midrule
\multirow{3}{*}{\textbf{GPT-Oss-20B}}
& \textit{(w/o) ranking-aware} & \underline{71.36} & \underline{87.26} & 175 & \underline{85.51} \\
& \textit{(w/o) model-aware} & 69.51 & 84.61 & 141 & 83.20 \\
& \textit{Ours} & \textbf{72.47} & \textbf{88.47} & 182 & \textbf{86.65} \\
\bottomrule
\end{tabular}
\label{tab:ablation}
\vspace{-0.3cm}
\end{table}

\subsubsection{Prompt-Sensitive Analysis.}
\begin{figure}[t]
    \setlength{\abovecaptionskip}{1pt}
    \centering
    \subfigure[Instance-level $\triangle$–$\triangle$ Analysis]{
        \centering
    \includegraphics[width=0.45\linewidth]{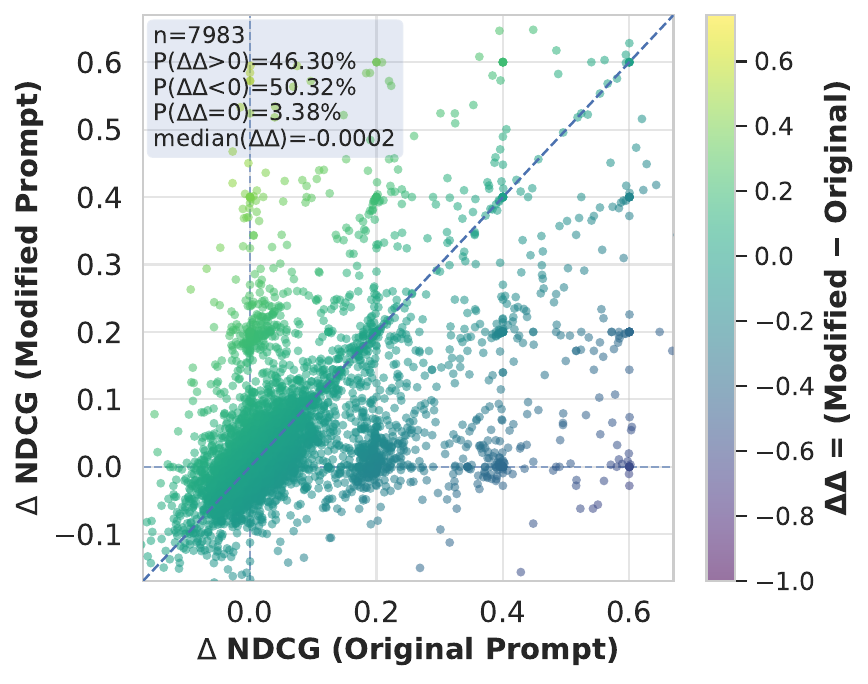}
        \label{fig:delta_delta_scatter}
    }
    \hspace{-0.2cm}
    \subfigure[Dataset-level Outcome Summary]{
        \centering
        \includegraphics[width=0.45\linewidth]{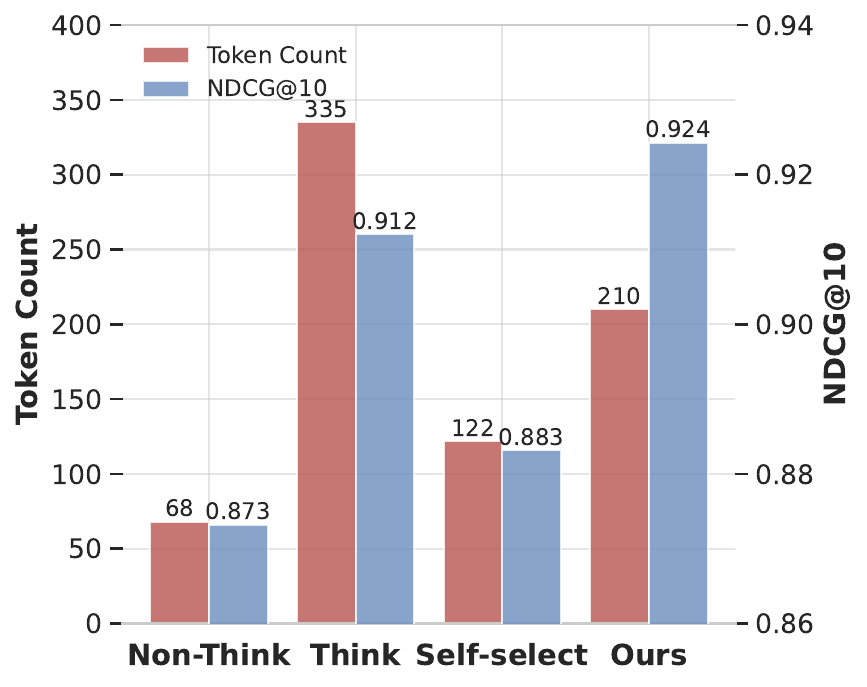}
      \label{fig:modified_prompt}
    }
      \caption{Prompt-sensitive analysis using Qwen3-14B on MS-MARCO.} 
  \label{fig:prompt-sensitive}
\end{figure}

To evaluate the robustness of reasoning routing under prompt perturbations, we conduct a prompt-sensitive analysis on MS-MARCO with Qwen3-14B. 
Starting from the original prompt used in Figure~\ref{fig:prompt_all_2}, we construct a modified prompt that adds explicit CoT steps, following the step-by-step reasoning template in \cite{sun2024large}.
We then repeat the same experimental procedure using the modified prompt, and quantify the instance-wise outcome shift induced by the prompt edit.
As shown in Figure~\ref{fig:delta_delta_scatter}, prompt modification can lead to substantial instance-level variability: the relative gains/losses in ranking utility fluctuate markedly across queries, indicating that \textit{\textbf{prompt phrasing meaningfully alters the effectiveness of reasoning on a per-instance basis}}. 

Despite this sensitivity, our method remains highly robust. 
Figure~\ref{fig:modified_prompt} shows that our router continues to achieve a favorable utility-efficiency trade-off under the modified prompt, delivering consistent improvements in ranking utility while reducing token consumption compared to baselines.
These results suggest that, \textbf{\textit{although prompt edits can change which instances benefit most from deeper reasoning, our routing mechanism is able to adaptively allocate computation and preserve its advantages}}, demonstrating robustness to realistic prompt engineering choices.

\section{Conclusion and Limitations}
This work introduced a \textit{lightweight, model-aware reasoning router} for LLM-based ranking tasks. 
By combining ranking-aware features and model-aware difficulty signals obtained through masked checklist probes, the proposed framework enables fine-grained, per-instance reasoning control without altering the backbone architecture. 
Extensive experiments across multiple datasets and LLM scales demonstrate that our approach consistently \textbf{improves ranking utility while substantially reducing token consumption}, achieving a better trade-off between efficiency and accuracy. 

Despite these encouraging results, our method has several limitations. 
First, the effectiveness of the router relies on a reasonable alignment between the training and test distributions of recommendation instances. 
When the test distribution shifts substantially in terms of user intent, candidate composition, or prompt format, the learned routing policy may become less reliable and, in the worst case, degenerate toward a nearly uniform reasoning mode. 
Second, our experiments focus on conversational recommendation scenarios, where the candidate set size is naturally constrained by the context length of LLM inputs. 
As a result, the current framework is mainly evaluated under relatively small candidate sets rather than large-scale retrieval or ranking pipelines. 
Future work will extend the proposed router to broader LLM-based recommendation settings, including retrieval-augmented ranking, multi-stage recommendation, and scenarios with larger candidate pools.

\section{Acknowledgments}
This research is supported by the State Key Laboratory of Industrial Control Technology, China (Grant No. ICT2026C02); the Ministry of Education, Singapore, under its Academic Research Fund (AcRF) Tier 1 grant, the SMU-SUTD Internal Research Grant Call (SMU-SUTD 2023\_02\_01); the Ministry of Education, Singapore, under its Academic Research Fund Tier 2 (Award No. MOE-T2EP201230015); and the MOE AcRF Tier 1 funding (RG13/23 and RG16/25).

\section*{Appendix}
\label{sec:appendix}
\subsection*{A. Checklist-based Probing (Pseudocode)}

Algorithm~\ref{alg:checklist-probing} summarizes the implementation of checklist-based probing.
For each ranking instance, we append a neutral probe and paired direction-balanced diagnostic questions to the shared ranking prefix.
A block-diagonal causal mask isolates different probe segments while allowing each probe to attend to the shared prefix.
In a single forward pass, hidden states at the \texttt{Answer:} anchors are mapped through the LM head to obtain \texttt{Yes}/\texttt{No} log-odds scores.
The neutral probe provides a baseline for reducing verbalizer bias.
The corrected probabilities and pairwise margins are then aggregated into model-aware difficulty signals and combined with ranking-aware features for router training.


\begin{algorithm}[h]
\footnotesize
\caption{Checklist-based probing mechanism.}
\label{alg:checklist-probing}
\DontPrintSemicolon
\LinesNumbered

\KwIn{
LLM $f_{\theta}$ with LM head $g_{\theta}$; tokenizer $\mathcal{T}$; ranking prompt $\pi=(P,C,\mathcal{I})$;\\
paired checklist $\mathcal{Q}=\{(q^{\mathrm{NT}}_j,q^{\mathrm{T}}_j)\}_{j=1}^{N_q}$
\tcp*{$\mathrm{NT}$ and $\mathrm{T}$ denote \nonthinkmode- and \thinkmode-oriented probes}
}
\KwOut{Checklist feature vector $\mathbf{x}_{\mathrm{ck}}$}

\SetKwFunction{EncodeChat}{EncodeChat}
\SetKwFunction{EncodeProbe}{EncodeProbe}
\SetKwFunction{BlockMask}{BlockDiagonalCausalMask}
\SetKwFunction{AnchorPos}{AnchorPos}
\SetKwFunction{Aggregate}{Aggregate}

$\mathbf{s} \leftarrow \EncodeChat(P,C,\mathcal{I})$
\tcp*{shared ranking prefix}

$\mathbf{r}_{0} \leftarrow \EncodeProbe(\texttt{``Answer: [PROBE]''})$
\tcp*{neutral probe}

\For{$j=1$ \KwTo $N_q$}{
    $\mathbf{r}^{\mathrm{NT}}_j \leftarrow
    \EncodeProbe(q^{\mathrm{NT}}_j,\texttt{``Answer: [PROBE]''})$\;

    $\mathbf{r}^{\mathrm{T}}_j \leftarrow
    \EncodeProbe(q^{\mathrm{T}}_j,\texttt{``Answer: [PROBE]''})$\;
}

$\mathbf{u} \leftarrow
[\mathbf{s};\mathbf{r}_{0};
\{\mathbf{r}^{\mathrm{NT}}_j,\mathbf{r}^{\mathrm{T}}_j\}_{j=1}^{N_q}]$\;

$\mathbf{M} \leftarrow
\BlockMask(\mathbf{s},\mathbf{r}_{0},
\{\mathbf{r}^{\mathrm{NT}}_j,\mathbf{r}^{\mathrm{T}}_j\}_{j=1}^{N_q})$
\tcp*{each probe attends only to $\mathbf{s}$ and itself}

$H \leftarrow f_{\theta}(\mathbf{u},\mathbf{M},\texttt{output\_hidden\_states}{=}\mathrm{True})$\;

$a_{0} \leftarrow \AnchorPos(\mathbf{r}_{0},\texttt{``Answer:''})$\;
$\beta \leftarrow
g_{\theta}(H_{a_0})_{v_{\texttt{Yes}}}
-
g_{\theta}(H_{a_0})_{v_{\texttt{No}}}$
\tcp*{neutral Yes/No log-odds}

\For{$j=1$ \KwTo $N_q$}{
    $a^{\mathrm{NT}}_j \leftarrow \AnchorPos(\mathbf{r}^{\mathrm{NT}}_j,\texttt{``Answer:''})$\;
    $a^{\mathrm{T}}_j \leftarrow \AnchorPos(\mathbf{r}^{\mathrm{T}}_j,\texttt{``Answer:''})$\;

    $r^{\mathrm{NT}}_j \leftarrow
    \big(
    g_{\theta}(H_{a^{\mathrm{NT}}_j})_{v_{\texttt{Yes}}}
    -
    g_{\theta}(H_{a^{\mathrm{NT}}_j})_{v_{\texttt{No}}}
    \big) - \beta$\;

    $r^{\mathrm{T}}_j \leftarrow
    \big(
    g_{\theta}(H_{a^{\mathrm{T}}_j})_{v_{\texttt{Yes}}}
    -
    g_{\theta}(H_{a^{\mathrm{T}}_j})_{v_{\texttt{No}}}
    \big) - \beta$\;

    $p^{\mathrm{NT}}_j \leftarrow \sigma(r^{\mathrm{NT}}_j)$;\quad
    $p^{\mathrm{T}}_j \leftarrow \sigma(r^{\mathrm{T}}_j)$\;

    $\Delta^{\mathrm{logit}}_j \leftarrow r^{\mathrm{NT}}_j-r^{\mathrm{T}}_j$;\quad
    $\Delta^{\mathrm{prob}}_j \leftarrow p^{\mathrm{NT}}_j-p^{\mathrm{T}}_j$\;
}

$\mathbf{x}_{\mathrm{ck}} \leftarrow
\Aggregate\left(
\{p^{\mathrm{NT}}_j,p^{\mathrm{T}}_j,
\Delta^{\mathrm{logit}}_j,\Delta^{\mathrm{prob}}_j\}_{j=1}^{N_q}
\right)$\;

\Return{$\mathbf{x}_{\mathrm{ck}}$}
\end{algorithm}

\newpage
\balance
\bibliographystyle{ACM-Reference-Format}
\bibliography{ref}

@inproceedings{cohen2023few,
  title={Few-sample feature selection via feature manifold learning},
  author={Cohen, David and Shnitzer, Tal and Kluger, Yuval and Talmon, Ronen},
  booktitle={International Conference on Machine Learning},
  pages={6296--6319},
  year={2023},
  organization={PMLR}
}

@article{madaan2023self,
  title={Self-refine: Iterative refinement with self-feedback},
  author={Madaan, Aman and Tandon, Niket and Gupta, Prakhar and Hallinan, Skyler and Gao, Luyu and Wiegreffe, Sarah and Alon, Uri and Dziri, Nouha and Prabhumoye, Shrimai and Yang, Yiming and others},
  journal={Advances in Neural Information Processing Systems},
  volume={36},
  pages={46534--46594},
  year={2023}
}

@inproceedings{yano2016quantifying,
  title={Quantifying query ambiguity with topic distributions},
  author={Yano, Yuki and Tagami, Yukihiro and Tajima, Akira},
  booktitle={Proceedings of the 25th ACM International on Conference on Information and Knowledge Management},
  pages={1877--1880},
  year={2016}
}

@article{wang2025kgbridge,
  title={KGBridge: Knowledge-Guided Prompt Learning for Non-overlapping Cross-Domain Recommendation},
  author={Wang, Yuhan and Xie, Qing and Bao, Zhifeng and Tang, Mengzi and Li, Lin and Liu, Yongjian},
  journal={arXiv preprint arXiv:2511.02181},
  year={2025}
}

@inproceedings{wang2025re2llm,
  title={Re2llm: reflective reinforcement large language model for session-based recommendation},
  author={Wang, Ziyan and Du, Yingpeng and Sun, Zhu and Chua, Haoyan and Feng, Kaidong and Wang, Wenya and Zhang, Jie},
  booktitle={Proceedings of the AAAI Conference on Artificial Intelligence},
  volume={39},
  number={12},
  pages={12827--12835},
  year={2025}
}

@article{achiam2023gpt,
  title={Gpt-4 technical report},
  author={Achiam, Josh and Adler, Steven and Agarwal, Sandhini and Ahmad, Lama and Akkaya, Ilge and Aleman, Florencia Leoni and Almeida, Diogo and Altenschmidt, Janko and Altman, Sam and Anadkat, Shyamal and others},
  journal={arXiv preprint arXiv:2303.08774},
  year={2023}
}

@article{li2025matching,
  title={From matching to generation: A survey on generative information retrieval},
  author={Li, Xiaoxi and Jin, Jiajie and Zhou, Yujia and Zhang, Yuyao and Zhang, Peitian and Zhu, Yutao and Dou, Zhicheng},
  journal={ACM Transactions on Information Systems},
  volume={43},
  number={3},
  pages={1--62},
  year={2025},
  publisher={ACM New York, NY}
}

@inproceedings{wang2025intent,
  title={Intent representation learning with large language model for recommendation},
  author={Wang, Yu and Sang, Lei and Zhang, Yi and Zhang, Yiwen},
  booktitle={Proceedings of the 48th International ACM SIGIR Conference on Research and Development in Information Retrieval},
  pages={1870--1879},
  year={2025}
}

@misc{openai2025gptoss120bgptoss20bmodel,
      title={gpt-oss-120b \& gpt-oss-20b Model Card}, 
      author={OpenAI},
      year={2025},
      eprint={2508.10925},
      archivePrefix={arXiv},
      primaryClass={cs.CL},
      url={https://arxiv.org/abs/2508.10925}, 
}

@article{gemma_2025,
    title={Gemma 3},
    url={https://goo.gle/Gemma3Report},
    publisher={Kaggle},
    author={Gemma Team},
    year={2025}
}

@inproceedings{sun-etal-2023-chatgpt,
    title = "Is {C}hat{GPT} Good at Search? Investigating Large Language Models as Re-Ranking Agents",
    author = "Sun, Weiwei  and
      Yan, Lingyong  and
      Ma, Xinyu  and
      Wang, Shuaiqiang  and
      Ren, Pengjie  and
      Chen, Zhumin  and
      Yin, Dawei  and
      Ren, Zhaochun",
    editor = "Bouamor, Houda  and
      Pino, Juan  and
      Bali, Kalika",
    booktitle = "Proceedings of the 2023 Conference on Empirical Methods in Natural Language Processing",
    month = dec,
    year = "2023",
    address = "Singapore",
    publisher = "Association for Computational Linguistics",
    url = "https://aclanthology.org/2023.emnlp-main.923/",
    doi = "10.18653/v1/2023.emnlp-main.923",
    pages = "14918--14937",
}

@inproceedings{liu2024information,
  title={Information retrieval meets large language models},
  author={Liu, Zheng and Zhou, Yujia and Zhu, Yutao and Lian, Jianxun and Li, Chaozhuo and Dou, Zhicheng and Lian, Defu and Nie, Jian-Yun},
  booktitle={Companion Proceedings of the ACM Web Conference 2024},
  pages={1586--1589},
  year={2024}
}

@inproceedings{hong2025llm,
  title={LLM-BS: Enhancing Large Language Models for Recommendation through Exogenous Behavior-Semantics Integration},
  author={Hong, Minjie and Xia, Yan and Wang, Zehan and Zhu, Jieming and Wang, Ye and Cai, Sihang and Yang, Xiaoda and Dai, Quanyu and Dong, Zhenhua and Zhang, Zhimeng and Zhou Zhao},
  booktitle={The Web Conference 2025}
}

@inproceedings{yin2025unleash,
  title={Unleash LLMs Potential for Sequential Recommendation by Coordinating Dual Dynamic Index Mechanism},
  author={Yin, Jun and Zeng, Zhengxin and Li, Mingzheng and Yan, Hao and Li, Chaozhuo and Han, Weihao and Zhang, Jianjin and Liu, Ruochen and Sun, Hao and Deng, Weiwei and others},
  booktitle={Proceedings of the ACM on Web Conference 2025},
  pages={216--227},
  year={2025}
}

@inproceedings{sun2024large,
  title={Large language models for intent-driven session recommendations},
  author={Sun, Zhu and Liu, Hongyang and Qu, Xinghua and Feng, Kaidong and Wang, Yan and Ong, Yew Soon},
  booktitle={Proceedings of the 47th International ACM SIGIR Conference on Research and Development in Information Retrieval},
  pages={324--334},
  year={2024}
}

@article{chen2024toward,
  title={Toward adaptive reasoning in large language models with thought rollback},
  author={Chen, Sijia and Li, Baochun},
  journal={arXiv preprint arXiv:2412.19707},
  year={2024}
}

@article{wei2022chain,
  title={Chain-of-thought prompting elicits reasoning in large language models},
  author={Wei, Jason and Wang, Xuezhi and Schuurmans, Dale and Bosma, Maarten and Xia, Fei and Chi, Ed and Le, Quoc V and Zhou, Denny and others},
  journal={Advances in Neural Information Processing Systems},
  volume={35},
  pages={24824--24837},
  year={2022}
}

@article{harper2015movielens,
  title={The movielens datasets: History and context},
  author={Harper, F Maxwell and Konstan, Joseph A},
  journal={Acm Transactions on Interactive Intelligent Systems (TIIS)},
  volume={5},
  number={4},
  pages={1--19},
  year={2015},
  publisher={ACM New York, NY, USA}
}

@inproceedings{ni2019justifying,
  title={Justifying recommendations using distantly-labeled reviews and fine-grained aspects},
  author={Ni, Jianmo and Li, Jiacheng and McAuley, Julian},
  booktitle={Proceedings of the Conference on Empirical Methods in Natural Language Processing and the 9th International Joint Conference on Natural Language Processing (EMNLP-IJCNLP)},
  pages={188--197},
  year={2019}
}

@article{pradeep2023rankzephyr,
  title={RankZephyr: Effective and Robust Zero-Shot Listwise Reranking is a Breeze!},
  author={Pradeep, Ronak and Sharifymoghaddam, Sahel and Lin, Jimmy},
  journal={arXiv preprint arXiv:2312.02724},
  year={2023}
}

@article{wu2025arm,
  title={ARM: Adaptive Reasoning Model},
  author={Wu, Siye and Xie, Jian and Zhang, Yikai and Chen, Aili and Zhang, Kai and Su, Yu and Xiao, Yanghua},
  journal={arXiv preprint arXiv:2505.20258},
  year={2025}
}

@inproceedings{weirocketeval,
  title={RocketEval: Efficient automated LLM evaluation via grading checklist},
  author={Wei, Tianjun and Wen, Wei and Qiao, Ruizhi and Sun, Xing and Ma, Jianghong},
  booktitle={The Thirteenth International Conference on Learning Representations},
  year={2025}
}

@article{zheng2023judging,
  title={Judging llm-as-a-judge with mt-bench and chatbot arena},
  author={Zheng, Lianmin and Chiang, Wei-Lin and Sheng, Ying and Zhuang, Siyuan and Wu, Zhanghao and Zhuang, Yonghao and Lin, Zi and Li, Zhuohan and Li, Dacheng and Xing, Eric and others},
  journal={Advances in Neural Information Processing Systems},
  volume={36},
  pages={46595--46623},
  year={2023}
}

@inproceedings{mohan2011web,
  title={Web-search ranking with initialized gradient boosted regression trees},
  author={Mohan, Ananth and Chen, Zheng and Weinberger, Kilian},
  booktitle={Proceedings of the Learning to Rank Challenge},
  pages={77--89},
  year={2011}
}

@article{wan2025adapthink,
  title={AdapThink: Adaptive Thinking Preferences for Reasoning Language Model},
  author={Wan, Xu and Wang, Wei and Xu, Wenyue and Yin, Wotao and Song, Jie and Sun, Mingyang},
  journal={arXiv preprint arXiv:2506.18237},
  year={2025}
}

@article{aggarwal2025l1,
  title={L1: Controlling how long a reasoning model thinks with reinforcement learning},
  author={Aggarwal, Pranjal and Welleck, Sean},
  journal={arXiv preprint arXiv:2503.04697},
  year={2025}
}

@article{fu2024efficient,
  title={Efficient llm scheduling by learning to rank},
  author={Fu, Yichao and Zhu, Siqi and Su, Runlong and Qiao, Aurick and Stoica, Ion and Zhang, Hao},
  journal={Advances in Neural Information Processing Systems},
  volume={37},
  pages={59006--59029},
  year={2024}
}

@article{parry2024top,
  title={Top-down partitioning for efficient list-wise ranking},
  author={Parry, Andrew and MacAvaney, Sean and Ganguly, Debasis},
  journal={arXiv preprint arXiv:2405.14589},
  year={2024}
}

@inproceedings{dong2025understand,
  title={Understand what LLM needs: Dual preference alignment for retrieval-augmented generation},
  author={Dong, Guanting and Zhu, Yutao and Zhang, Chenghao and Wang, Zechen and Wen, Ji-Rong and Dou, Zhicheng},
  booktitle={Proceedings of the ACM on Web Conference 2025},
  pages={4206--4225},
  year={2025}
}

@inproceedings{sudualformer,
  title={Dualformer: Controllable Fast and Slow Thinking by Learning with Randomized Reasoning Traces},
  author={Su, DiJia and Sukhbaatar, Sainbayar and Rabbat, Michael and Tian, Yuandong and Zheng, Qinqing},
  booktitle={The Thirteenth International Conference on Learning Representations},
  year={2024}
}

@article{hao2024training,
  title={Training large language models to reason in a continuous latent space},
  author={Hao, Shibo and Sukhbaatar, Sainbayar and Su, DiJia and Li, Xian and Hu, Zhiting and Weston, Jason and Tian, Yuandong},
  journal={arXiv preprint arXiv:2412.06769},
  year={2024}
}

@article{hou2025thinkprune,
  title={Thinkprune: Pruning long chain-of-thought of llms via reinforcement learning},
  author={Hou, Bairu and Zhang, Yang and Ji, Jiabao and Liu, Yujian and Qian, Kaizhi and Andreas, Jacob and Chang, Shiyu},
  journal={arXiv preprint arXiv:2504.01296},
  year={2025}
}

@article{tang2025think,
  title={Think before recommend: Unleashing the latent reasoning power for sequential recommendation},
  author={Tang, Jiakai and Dai, Sunhao and Shi, Teng and Xu, Jun and Chen, Xu and Chen, Wen and Wu, Jian and Jiang, Yuning},
  journal={arXiv preprint arXiv:2503.22675},
  year={2025}
}

@article{pereira2024check,
  title={Check-Eval: A checklist-based approach for evaluating text quality},
  author={Pereira, Jayr and Assumpcao, Andre and Lotufo, Roberto},
  journal={arXiv preprint arXiv:2407.14467},
  year={2024}
}

@article{jarvelin2002cumulated,
  title={Cumulated gain-based evaluation of IR techniques},
  author={J{\"a}rvelin, Kalervo and Kek{\"a}l{\"a}inen, Jaana},
  journal={ACM Transactions on Information Systems (TOIS)},
  volume={20},
  number={4},
  pages={422--446},
  year={2002},
  publisher={ACM New York, NY, USA}
}

@inproceedings{zhang2025hierarchical,
  title={Hierarchical Time-Aware Mixture of Experts for Multi-Modal Sequential Recommendation},
  author={Zhang, Shengzhe and Chen, Liyi and Shen, Dazhong and Wang, Chao and Xiong, Hui},
  booktitle={Proceedings of the ACM on Web Conference 2025},
  pages={3672--3682},
  year={2025}
}

@inproceedings{wang2025enhancing,
  title={Enhancing Transferability and Consistency in Cross-Domain Recommendations via Supervised Disentanglement},
  author={Wang, Yuhan and Xie, Qing and Bao, Zhifeng and Tang, Mengzi and Li, Lin and Liu, Yongjian},
  booktitle={Proceedings of the Nineteenth ACM Conference on Recommender Systems},
  pages={104--113},
  year={2025}
}

@inproceedings{beutel2019fairness,
  title={Fairness in recommendation ranking through pairwise comparisons},
  author={Beutel, Alex and Chen, Jilin and Doshi, Tulsee and Qian, Hai and Wei, Li and Wu, Yi and Heldt, Lukasz and Zhao, Zhe and Hong, Lichan and Chi, Ed H and others},
  booktitle={Proceedings of the 25th ACM SIGKDD International Conference on Knowledge Discovery \& Data Mining},
  pages={2212--2220},
  year={2019}
}

@article{ekstrom2019sequential,
  title={Sequential rank agreement methods for comparison of ranked lists},
  author={Ekstr{\o}m, Claus Thorn and Gerds, Thomas Alexander and Jensen, Andreas Kryger},
  journal={Biostatistics},
  volume={20},
  number={4},
  pages={582--598},
  year={2019},
  publisher={Oxford University Press}
}

@misc{qwen3technicalreport,
      title={Qwen3 Technical Report}, 
      author={Qwen Team},
      year={2025},
      eprint={2505.09388},
      archivePrefix={arXiv},
      primaryClass={cs.CL},
      url={https://arxiv.org/abs/2505.09388}, 
}

@inproceedings{shah2016pareto,
  title={Pareto frontier learning with expensive correlated objectives},
  author={Shah, Amar and Ghahramani, Zoubin},
  booktitle={International Conference on Machine Learning},
  pages={1919--1927},
  year={2016},
  organization={PMLR}
}

@inproceedings{jiang2023llm,
  title={LLM-Blender: Ensembling Large Language Models with Pairwise Ranking and Generative Fusion},
  author={Jiang, Dongfu and Ren, Xiang and Lin, Bill Yuchen},
  booktitle={Proceedings of the 61st Annual Meeting of the Association for Computational Linguistics (Volume 1: Long Papers)},
  pages={14165--14178},
  year={2023}
}

@inproceedings{podolak2025beyond,
  title={Beyond reproducibility: Advancing zero-shot llm reranking efficiency with setwise insertion},
  author={Podolak, Jakub and Peri{\'c}, Leon and Jani{\'c}ijevi{\'c}, Mina and Petcu, Roxana},
  booktitle={Proceedings of the 48th International ACM SIGIR Conference on Research and Development in Information Retrieval},
  pages={3205--3213},
  year={2025}
}

@inproceedings{zhuang2024setwise,
  title={A setwise approach for effective and highly efficient zero-shot ranking with large language models},
  author={Zhuang, Shengyao and Zhuang, Honglei and Koopman, Bevan and Zuccon, Guido},
  booktitle={Proceedings of the 47th International ACM SIGIR Conference on Research and Development in Information Retrieval},
  pages={38--47},
  year={2024}
}

@article{liu2025reasonrank,
  title={ReasonRank: Empowering Passage Ranking with Strong Reasoning Ability},
  author={Liu, Wenhan and Ma, Xinyu and Sun, Weiwei and Zhu, Yutao and Li, Yuchen and Yin, Dawei and Dou, Zhicheng},
  journal={arXiv preprint arXiv:2508.07050},
  year={2025}
}

@inproceedings{guo2024configurable,
  title={Configurable fairness for new item recommendation considering entry time of items},
  author={Guo, Huizhong and Wang, Dongxia and Sun, Zhu and Zhang, Haonan and Li, Jinfeng and Zhang, Jie},
  booktitle={Proceedings of the 47th International ACM SIGIR Conference on Research and Development in Information Retrieval},
  pages={437--447},
  year={2024}
}

\end{document}